\documentclass[letterpaper]{article} 
\usepackage{aaai2026}  
\usepackage{times}  
\usepackage{helvet}  
\usepackage{courier}  
\usepackage[hyphens]{url}  
\usepackage{graphicx} 
\usepackage{natbib}  
\usepackage{caption} 
\usepackage{algorithm}
\usepackage{algorithmic}

\usepackage{newfloat}
\usepackage{listings}
\DeclareCaptionStyle{ruled}{labelfont=normalfont,labelsep=colon,strut=off} 
\floatstyle{ruled}
\newfloat{listing}{tb}{lst}{}
\floatname{listing}{Listing}
\usepackage{dirtytalk}
\usepackage{multirow}
\usepackage{amsmath}
\usepackage{amssymb}
\usepackage{amsfonts}
\usepackage{makecell}
\usepackage{bm} 
\usepackage{scalerel}
\newlength\bshft
\def\fakebold#1{\ThisStyle{\ooalign{$\SavedStyle#1$\cr%
  \kern-\bshft$\SavedStyle#1$\cr%
  \kern\bshft$\SavedStyle#1$}}}

\usepackage{xcolor}

\title{Meme Template Identification in the Wild:\\Comparing Methods for Semi-Open-Set Recognition
}
\author{
    H. Nguyen\textsuperscript{\rm 1}\equalcontrib, D. Á. Köller\textsuperscript{\rm 2}\equalcontrib, L. Murgás\textsuperscript{\rm 2}, J. Pintér\textsuperscript{\rm 2},  M. Nagy\textsuperscript{\rm 2}, K. Barnes\textsuperscript{\rm 3}, R. Molontay\textsuperscript{\rm 2}
}
\affiliations{

    \textsuperscript{\rm 1}Department of Math and Computer Science, Augustana College,
    639 38th St, Rock Island, IL, 61201, USA\\
    \textsuperscript{\rm 2}Department of Stochastics, Budapest University of Technology and Economics,\\
    Műegyetem rakpart 3., 1111 Budapest, Hungary\\
    \textsuperscript{\rm 3}University of Colorado Boulder, 40 UCB, Boulder, Colorado 80309-0040, USA\\

}

\begin{document}

\maketitle

\begin{abstract}
Image-with-text memes are a dominant form of online communication, and much of their spread happens through meme templates which are recurring visual formats that users adapt with new text or imagery. Most prior work scores memes individually for engagement or harmful content, an approach that is structurally blind to templates. Templates might amplify these patterns and enable coordinated harassment, which becomes visible only when memes are grouped by shared template. We formalize meme template identification as a semi-open-set recognition problem, requiring methods to both classify known templates and reject template-free memes and non-meme content. We introduce an evaluation framework spanning a controlled setting (1,704 ImgFlip templates) and a heterogeneous real-world social media sample, comparing supervised CNN- and distance-based methods, unsupervised density-based clustering, a novel fused SigLIP2+DINOv2 representation, and a retrieval-augmented LLM pipeline. In real-world social media sources dedicated to meme sharing, we find that only 21\% of images use a known template, while 58\% are template-free and 21\% are not memes at all; identification performance also drops sharply from the controlled setting (best MCC 0.974 to 0.684). The loss comes mainly from deciding whether an image is a template-based meme at all, rather than identifying the template. All data are archived on Zenodo, including 1,117,847 social media images with template labels predicted by GPT-5.6 Luna, the 152,509 ImgFlip images with their three train/test partitions, and the annotated 1,000-image benchmark. The code, which reads this archive directly, is available online to facilitate reproducibility.

\end{abstract}

\begin{links}
    \link{Code}{https://anonymous.4open.science/r/CodeAndDataSupplement/README.md}
    \link{Datasets}{https://doi.org/10.5281/zenodo.23676782}
\end{links}

\section{Introduction}
In the contemporary digital landscape, user attention gravitates towards easily digestible media like short-form videos and, notably, image-with-text memes. A meme's power lies in its multimodal nature, compressing a wealth of information (including cultural references, humor, and political commentary) into a single, shareable artifact that combines textual and visual elements \cite{davison2012language}. Far from being mere digital amusement, memes have evolved into a potent medium, exerting significant influence across societal domains, from manipulating political dialogues to marketing strategies \cite{leach2017social, malodia2022meme, shifman2013memes, wang2025internet}.\footnote{The meme examples presented in this paper are sourced from public online platforms and are included solely for illustrative and analytical purposes. Their inclusion does not constitute an endorsement of the views or opinions they express, which belong to their original creators. Reader discretion is advised as some content may be considered sensitive or offensive.}

Template-level analysis has concrete downstream uses: content moderation teams need to detect coordinated inauthentic behavior, where a template is deployed at scale to manufacture organic-looking sentiment, and content reuse is already an established coordination-detection signal \cite{luceri2026coordinated}; researchers studying disinformation similarly need to trace how narratives are adapted across communities by template, not merely by topic or hashtag \cite{qu2022disinfomeme}.

To understand these phenomena at scale, we must analyze memes not just as individual instances but through their shared foundations. Central to this is the concept of a \textbf{meme template}: the structural and semantic backbone that allows for a wide range of variations while being grounded in a recognizable format \cite{hazman2025makes, bates2023template}, most often realized as \say{Character Macros} or \say{Format Macros} (formalized, along with the template-based/template-free boundary, in Related Work). Different meme instances can derive from the same underlying template while varying in overlaid text or imagery (e.g., from the well-known \say{Distracted Boyfriend} format; see Figure \ref{fig:template} in the Appendix, \textit{Template Example}). Identifying these templates is a crucial first step for tracking the spread of ideas and narratives at a higher level of abstraction, facilitating studies on meme evolution, virality, and influence on public discourse.

\begin{figure}[!ht]
    \centering
    \includegraphics[width=\columnwidth]{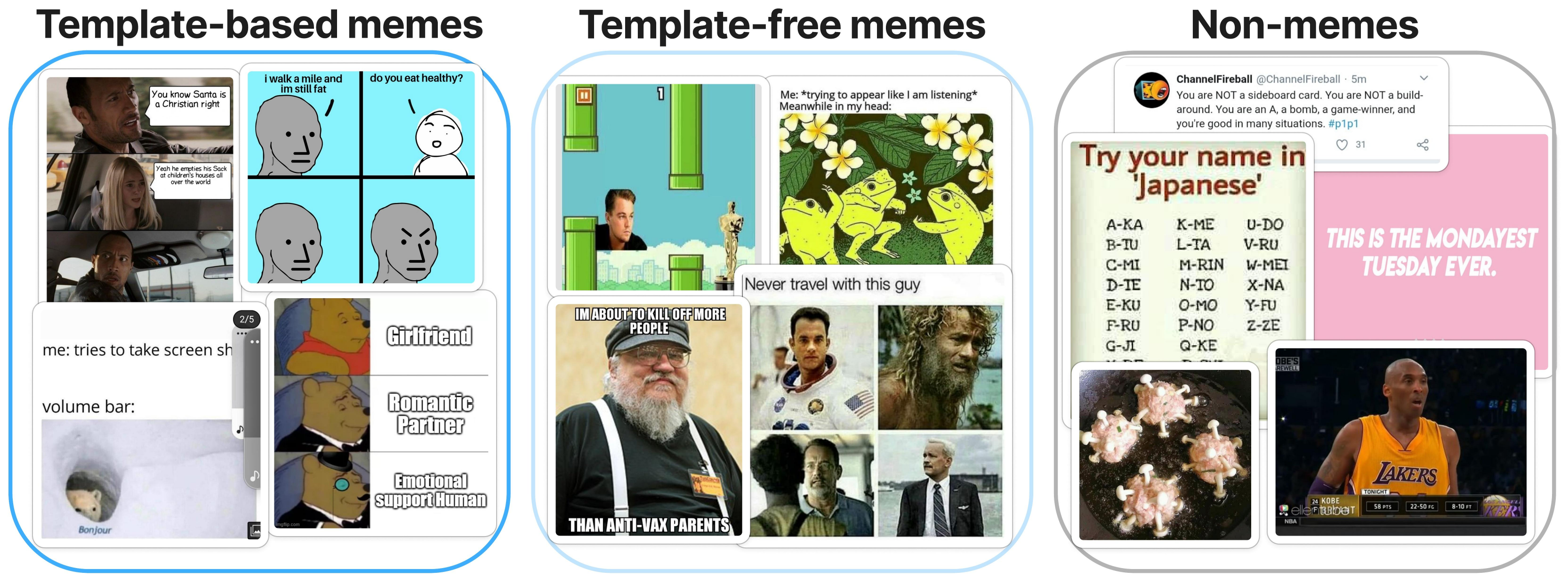}
    \caption{Examples of the template-based memes, template-free memes, and non-memes from the social media dataset.}
\label{fig:templated-vs-non-templated}
\end{figure}

However, the \say{wild} of social media is not composed solely of clean, template-based memes: even sources explicitly dedicated to meme-sharing (subreddits, Facebook groups, hashtag-curated tweet streams) yield a heterogeneous mix of \textbf{template-based memes} (an established format), \textbf{template-free memes} (one-off memetic creations with no recurring template), and \textbf{non-meme images} (incidental visual content with no memetic intent), illustrated in Figure \ref{fig:templated-vs-non-templated} and defined fully in Related Work. Our own annotated sample of social media images (\textit{Evaluation Setting 2}) provides empirical evidence of this skew: only 21\% of images were template-based, against 58.3\% template-free and 20.7\% non-meme.

Most existing identification methods are developed and validated only on clean, closed-set collections such as curated template libraries \cite{dubey2018memesequencer, courtois2023computer, joshi2024contextualizing, sherratt2026clustering}. A method that performs well there but cannot reliably reject template-free and non-meme content is not fit for the applications motivating this work. Our setting is instead semi-open-set: we seek to assign templates to unlabeled (memetic) images collected from the \say{open wilderness} of social media sources dedicated to meme sharing, while assuming access to a closed list of known templates that can be used for model training. To systematically investigate this issue, we address the following research questions (RQs):

\begin{enumerate}
\item[\textbf{RQ1:}] \textbf{Performance in a Controlled Setting (meme template classification)} How effectively do different supervised and unsupervised methods perform at classifying known meme templates when evaluated on a large, clean dataset composed solely of template-based memes (i.e., ImgFlip)?

\item[\textbf{RQ2:}] \textbf{Performance in a Real-World Setting (meme template identification)} How robust are these models in a real-world scenario when confronted with a heterogeneous mix of template-based, template-free, and non-meme images? Specifically, how well do they distinguish known templates from the broader category of \say{template-free} content?

\item[\textbf{RQ3:}] \textbf{Efficiency and Scalability} What are the practical trade-offs between prediction performance, time efficiency, and scalability across the various identification methods when applied to large-scale datasets?

\end{enumerate}


\begin{figure}[t]
    \centering
    \includegraphics[width=\columnwidth]{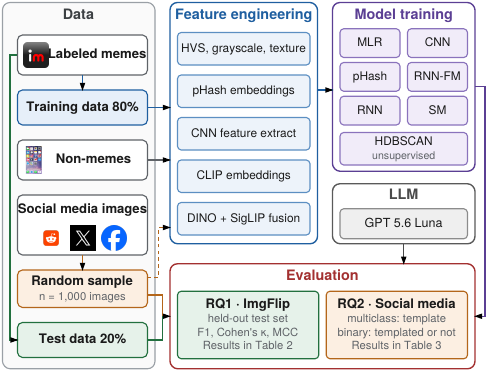} 
    \caption{Overall workflow of our research.}
    \label{fig:framework}
\end{figure}

To answer these questions, this paper combines a structured, labeled source (ImgFlip) with unstructured, unlabeled social media data (Reddit, Facebook, $\mathbb{X}$), and implements, trains, and evaluates a wide array of supervised and unsupervised methods against an evaluation framework that scores not only correct template classification but also correct rejection of non-template-based content (Figure \ref{fig:framework}). Because a meme template is most reliably carried by its persistent visual layout—background image, panel structure, character pose—while overlaid text is the locus of user customization and is frequently absent, OCR-degraded, or multilingual, we scope this study to vision-only identification.

The main contributions of this paper are threefold. First and foremost, we formalize meme template identification as a semi-open-set recognition problem rather than a closed-set classification task, and provide the first comprehensive framework for comparing vision-only identification methods across a controlled setting (RQ1), a real-world heterogeneous setting (RQ2), and efficiency/scalability (RQ3). Second, within this framework, we additionally propose and evaluate two novel entrants alongside established literature baselines: a fused SigLIP2 and DINOv2 embedding that combines language-aligned semantic similarity with layout-sensitive visual structure, and a retrieval-augmented LLM pipeline that prunes the template inventory to a small candidate set before verification, avoiding the need to present the full label space to the model at inference time. Third, we provide an extensive dataset of 1.1 million social media images, soft-labeled with the template and meme/non-meme predictions produced by the methods evaluated in this paper, along with code to support reproducibility and follow-up work.

\section{Related Work}

\subsection{Defining Memes and Meme Templates}

Before situating our work relative to prior template-identification methods, we clarify the conceptual object of study, since "meme" is used inconsistently across the literature this paper draws on.

Following Dawkins' original formulation, as popularized for the internet age, a meme is a unit of culture (an idea, behavior, or style) that spreads through imitation and variation \cite{davison2012language}. Shifman's influential account further identifies memes as characterized by three dimensions along which imitation and variation occur: content, form, and stance \cite{shifman2013memes}. Image-with-text memes are a specific, highly visible instantiation of this broader concept, in which imitation is realized through a shared form, most often a recognizable image or layout, while content and stance vary across instances. It is this shared form that we operationalize as a "template."

We define a \textbf{meme template} as a recurring visual form that is reused across multiple independently created instances, with each instance introducing new textual or visual content while preserving the recognizable form. This visual unit can take the form of an image, layout, or combination of the two. Our definition of a meme template has two consequences that matter for our task. First, \say{recurring} is doing load-bearing work: templatehood is not an intrinsic property of an image but a property of its usage pattern across a population of memes. An image is not a template because of what it depicts, but because a community has, in practice, reused it. Second, this means template status can only be assessed relative to an observed corpus, which is precisely why we distinguish between the closed-set setting (RQ1), where templatehood is given by construction, and the semi-open-set setting (RQ2), where it must be inferred from a heterogeneous and only partially observed stream of content. In both settings, images are compared to a previously established set of templates (ImgFlip in our case).

Following the typology of \citet{hazman2025makes}, we further distinguish two common sub-forms of template-based memes: \textit{Character Macros}, where text is overlaid on a fixed background image (e.g., \say{Distracted Boyfriend}; see Figure~\ref{fig:template} in the Appendix), and \textit{Format Macros}, where a fixed spatial layout of image-and-text regions is reused with fully substitutable content (e.g., multi-panel comparison formats). Both are covered by our definition above and by our dataset, since ImgFlip's template library spans both sub-forms.

\textit{The ambiguous middle case: Memetic Images}. A more difficult case, and one central to the difficulty of semi-open-set identification, is what \citet{hazman2025makes} term \textit{Memetic Images}: posts in which a captioning style or textual joke format is applied to an image that is not itself drawn from a fixed, recurring visual pool. Critically, whether a given Memetic Image instance counts as template-based or template-free cannot be determined from the image alone; it depends on whether that specific background image recurs across many independently authored instances. Here, we operationalize \say{recurring} relative to the observed ImgFlip corpus rather than as an unverifiable claim about all possible future or unseen usage. Concretely, an image is treated as template-based if (a) it appears in ImgFlip's curated template library, where recurrence is guaranteed by construction, or (b) for the evaluation on social media content without such ground truth, it can be matched by a human annotator using reverse image search and platform search tools to an existing template found in our ImgFlip library. Memetic content that cannot be matched to any of the templates in the ImgFlip library is labeled template-free, with the explicit caveat that this is a corpus-relative, not ontological, judgment: an image labeled template-free may later recur and become template-based.

For example, Figure \ref{fig:templated-vs-non-templated} places two single-image-with-caption memes side by side that are, at the level of visual layout, indistinguishable: the template-based example (\say{Bonjour Polar Bear}, bottom-left) and the template-free example (\say{4 panels Tom Hanks}, bottom-right) both consist of one image with a short caption above it, differing only in which specific image is used. What separates them is not visible in either image alone: it is that the first recurs across many independently authored ImgFlip instances and the second, as far as our corpus shows, does not. We treat this ambiguity as a defining challenge of the identification task rather than a labeling nuisance: it is precisely why our evaluation framework treats template membership as something a model must infer from population-level evidence (recurrence), not from single-image content or layout alone. This same population-level, recurrence-based reasoning is also what motivates our decision to scope this study to visual templates and exclude purely textual \textit{Memetic Trends}, where no visual background recurs at all. We return to this scoping decision below (Research Gap).

A stricter memetics tradition treats replication as constitutive of memehood, so a singleton, unreplicated image would not count as a meme at all \cite{shifman2013memes, Milner2016, davison2012language}. We depart from this for two reasons: replication is unobservable at collection time (an item may simply not yet have recurred), and collapsing non-recurring content into a non-meme category would remove the hard negative class our evaluation needs \cite{sherratt2023multi}. We therefore retain \say{template-free meme} as a description of visual form, not a claim about replication history.

\textit{Non-memes}. Finally, we distinguish both categories above from \textbf{non-meme images}: visual content that is neither template-based nor memetic in intent (e.g., incidental screenshots, personal photos, or motivational quote graphics posted in meme-sharing spaces without humorous or memetic function). These form the second component of the rejection class alongside template-free memes (see Data Collection).

To ground these categories operationally rather than leaving them as abstract definitions, our human annotators (see Evaluation Setting 2) applied the following criterion when labeling the 1,000-image social media sample: an image was labeled template-based if the meme-expert annotator could identify, or reasonably infer from visual conventions, that its background/layout recurs across independently authored posts; template-free if it appeared to be a one-off memetic creation; and non-meme if it showed no evidence of memetic intent. We report inter-annotator agreement for this labeling scheme in Evaluation Setting 2.

\subsection{Meme vs. Non-Meme Classification}

A related but distinct line of work asks a coarser question than ours: not which template an image belongs to, but whether it is a meme at all. \citet{Beskow2020} fuse image and text features to separate memes from non-memes and trace meme families genealogically via image similarity and OCR, a framing close in spirit to our own tracking motivation, but one that stops at the meme/non-meme boundary rather than resolving template identity. \citet{Du2020} build a comparable image-with-text (IWT) vs. general-image classifier, falling back on the pipeline of \citet{zannettou2018origins} to separate out non-IWT memes, illustrating how this literature treats meme identification and non-meme rejection as separate sub-problems. \citet{SharmaPulaBaigariDas2020} instead use a Siamese network to model the visuo-linguistic coupling between an image and its overlaid text, on the premise that this coupling is tighter in genuine memes than in incidental IWT content.

Two further studies interrogate what makes a good negative class for meme detection, directly informing our own Non-Meme dataset design (see Data Collection): \citet{Koutlis2023} show that classifiers trained on generic-photograph negatives learn superficial heuristics and fail against harder negatives drawn from other IWT formats, while \citet{sherratt2023multi} make a closely related argument with a multi-channel CNN trained against a deliberately IWT-blended negative class.

Collectively, these five studies establish that meme-vs-non-meme classification is non-trivial once negatives are drawn from other IWT formats rather than generic photographs, and that cues like text presence alone are insufficient for robust detection. However, none of these studies distinguishes template-based from template-free memes on the positive side of that boundary: any image exhibiting memetic form or stance is simply "a meme," regardless of whether its visual background recurs across many independently authored instances or was used exactly once. Our task requires going one step further: because our goal is to recover which template an image was generated from, we must additionally separate template-free memes, which would pass every meme-vs-non-meme test in this literature, from template-based ones, for which a specific answer exists. Our semi-open-set evaluation (RQ2) therefore subsumes this binary task, pooling template-free memes and non-memes into a single \say{non-template-based} rejection category and requiring a strictly harder within-meme discrimination that these works do not address.

\subsection{Prior Approaches to Template Identification}

The task of automatically identifying and analyzing meme templates has been approached from several perspectives in prior research. Early foundational work focused on tracking the spread and evolution of memes across platforms: \citet{zannettou2018origins} introduced a pipeline using perceptual hashing (pHash) and DBSCAN clustering to track memes across multiple web communities, establishing a strong baseline for image-similarity-based grouping, though relying on post-hoc annotation without explicitly modeling non-meme rejection. Subsequent work moved toward richer feature representations: \citet{dubey2018memesequencer} proposed sparse representation to decouple the \say{local context} (overlaid text/images) from the \say{global context} (the template itself), enabling multimodal features from both components; \citet{courtois2023computer} combined keypoint feature matching with network analysis, introducing a text-blurring preprocessing step to isolate visual features from text; and \citet{theisen2023motif} introduced \say{motif mining}, combining global and local image features to find remixed content in large, unlabeled collections.

Other works approach the problem from a knowledge organization perspective. \citet{tommasini2023imkg} introduced the Internet Meme Knowledge Graph (IMKG), a rich repository for representing memes and templates; building on this, \citet{joshi2024contextualizing} developed a framework to contextualize social media memes by mapping them to IMKG using vision transformer-based similarity. While their goal of contextualization is similar to our task, their approach relies on matching social media images against templates cataloged in IMKG, which are primarily sourced from \say{Know Your Meme: Internet Meme Database} (KYM) and, as we detail in Data Collection, inconsistently curated. This limitation is independently corroborated by \citet{sherratt2026clustering}, whose metric-learning pipeline for aligning memes to KYM categories documents the same submission-approval inconsistency. We instead build our ground-truth set from ImgFlip, where the meme-to-template link is programmatically guaranteed.

A particularly relevant recent work by \citet{hazman2025makes} develops a detailed typology for multimodal memes grounded in memetics, categorizing memes into \say{Template Memes}, \say{Memetic Images}, \say{Transferred Symbols}, and \say{Memetic Trends}, and further subdividing \say{Template Memes} into Character and Format Macros. Our work aligns with and extends this perspective. While we adopt a similar understanding of template-based formats, our research highlights a critical ambiguity that must be addressed for robust classification: formats like \say{Memetic Images} can function as either template-based or template-free, depending on whether the underlying image is reused across many instances or captioned once, as illustrated in Figure \ref{fig:templated-vs-non-templated}. This distinction is central to our objective of creating models that can handle the complexities of real-world data where such ambiguities are common. Our study focuses on the categories where the template is visual, leaving the identification of text-based formats like \textit{Memetic Trends} as an important avenue for future multimodal research.

Beyond the meme-vs-non-meme literature above, a broader body of work studies multimodal \textit{understanding} of meme content (humor, sarcasm, stance, harm) typically by aligning visual and textual encoders so the image–caption relationship itself carries the predicted label. \citet{liang2026computational} survey this space and note that multimodal alignment classifies \textit{what} a meme communicates while the underlying mechanism remains largely implicit. Template identification asks a different question: whether an image's visual form recurs, a property of the image rather than the caption (see \textit{Defining Memes and Meme Templates}), which motivates our vision-only scope.

\paragraph{Research Gap}
While these studies provide crucial foundations, they share a common limitation: their evaluation is often confined to datasets primarily consisting of known, template-based memes, without explicitly accounting for the significant presence of template-free memes and non-meme images that populate real-world social media feeds originally designed for sharing memes. Our work directly addresses this gap by establishing a framework where the ability to reject these out-of-distribution categories is a primary evaluation criterion.

We additionally scope this comparison to \textbf{vision-only} identification, consistent with the distinction drawn above; the fuller technical rationale is in Methodology.

While a separate literature has established that robust meme-vs-non-meme classification requires hard negatives drawn from other IWT formats rather than generic photographs \cite{Beskow2020, Du2020, SharmaPulaBaigariDas2020, Koutlis2023, sherratt2023multi}, none of these works distinguish template-free from template-based memes within the positive class, which is a distinction that is orthogonal to, and unaddressed by, the meme/non-meme boundary they study.

\section{Data Collection}\label{data-collection}

\begin{table*}[t]
\centering
\small
\begin{tabular}{l|l|r|l}
\hline
\textbf{Type}              & \textbf{Data Source}                         & \textbf{No. of Samples} & \textbf{No. of Templates} \\ \hline
\makecell[l]{\textbf{Template-based memes} \\ {\footnotesize(with ground-truth template labels)}}                     & \makecell[l]{ImgFlip \\ {(\url{imgflip.com/memetemplates})}}    & 152,509\,\,\,\,\,\,              & 1,704                  \\ \hline
\multirow{3}{*}{\makecell[l]{\textbf{Memes and non-memes} \\ {\footnotesize(template-based memes, template-free memes, and non-} \\{\footnotesize  memes in unknown proportions without labels)}}}     & Reddit (\texttt{r/Memes})                    & 707,137\,\,\,\,\,\,              & unlabeled              \\
                           & Facebook (meme-sharing groups)               & 236,371\,\,\,\,\,\,              & unlabeled              \\
                           & $\mathbb{X}$ (tweets with \#meme or \#dank) & 174,339\,\,\,\,\,\,              & unlabeled              \\ \hline
\multirow{4}{*}{\textbf{Non-memes}} & Screenshots \cite{deka2017rico}                             & 66,261\,\,\,\,\,\,               & unlabeled              \\
                            & Flickr30k \cite{plummer2015flickr30k}                       & 31,783\,\,\,\,\,\,               & unlabeled              
                           \\
                           & Movie posters                                                        & 8,252\,\,\,\,\,\,               & unlabeled              \\ \hline
\end{tabular}
\caption{Overview of datasets. The proportion of memes and non-memes in the social media data is unknown.}
\label{tab:datasets}
\end{table*}

Data collection served two purposes: ImgFlip provided the core training dataset with reliable ground-truth template labels, while Reddit, $\mathbb{X}$, and Facebook provided real-world, unlabeled data, serving as a discovery corpus for the unsupervised methods and, via a 1,000-image human-annotated sample, for semi-open-set evaluation. The Flickr30k and Screenshots collections instead supplied non-meme training samples for the dataset (see \textit{Non-memes}, below).

\paragraph{ImgFlip.}
ImgFlip is a popular website for creating and sharing image-with-text and GIF memes. The platform's structure as a meme generator is particularly advantageous for our work: when users create memes, they are inherently linked to the specific template used, thus providing a readily available and explicit label (the template) for each meme instance. We considered KYM as an alternative source, but its \say{example} images for meme entries are not consistently curated to be direct products of the associated template. Often, these examples might only share a thematic connection or could be instances where a tag was applied loosely, rather than the image being a direct generation from that template. This lack of reliably accurate template-to-instance pairing makes KYM's data unsuitable for training machine learning algorithms that depend on such precise labeling. Therefore, ImgFlip’s system of generating memes from specific templates provided the necessary data integrity for our study.

We build on the ImgFlip portion of the dataset assembled by \citet{lememematch2026}. To search for additional possible duplicates, we performed a screening using a combination of CLIP-embedding similarity and LLM-flagged candidates, with every proposed merge manually reviewed by the authors (full procedure in the Appendix, \textit{ImgFlip Deduplication}). After that, we were left with 1,770 canonical template identities, comprising 152,767 memes. We then required at least seven images per template so that every class could be split across training and test partitions; this removed 66 sparsely populated templates holding 258 images in total, leaving a dataset of 152,509 memes across 1,704 canonical templates. A canonical class holds 89.5 memes on average (median: 87), with class sizes ranging from 7 to 761.

\paragraph{Social Media Data (Reddit, $\mathbb{X}$, Facebook).}
To assess model generalization on real-world data, we collected images from three major platforms.
\textbf{Reddit:} We gathered 707,137 images from the \texttt{r/memes} subreddit posted between Jan 2018 and Dec 2023 using the Pushshift API.
\textbf{$\mathbb{X}$ (formerly Twitter):} Using Selenium, we collected 174,339 images from tweets containing the hashtags \say{dank} or \say{meme} published between Jan 2018 and Jun 2020.
\textbf{Facebook:} We scraped 20 English-speaking meme-sharing groups from Sept 2013 to Oct 2023, yielding 236,371 images.
Importantly, for all platforms, we targeted only shared post content and excluded user profile pictures or sensitive personal data.

\paragraph{Non-memes.}
Although our data collection targeted online communities dedicated to sharing memes, the process inevitably captured template-free memes and non-meme images, as defined in Related Work (e.g., motivational quote graphics, screenshots, and landscape photos with no memetic intent).

To effectively distinguish template-based memes from these other image types, we curated a dedicated \say{Non-Meme} dataset following the hard-negative design principles established in Related Work (\textit{Meme vs. Non-Meme Classification}): an equal mixture of Flickr30k photographs and other IWT formats (movie posters, screenshots). Classifiers trained only against generic photographs tend to learn superficial heuristics (e.g., mere text presence) that fail against harder IWT-format negatives; in our observations, these harder negatives also more closely resemble incidental social media content than the distinctly structured template-based memes on ImgFlip do.

The number of images and posts collected from the various online platforms and the Non-Meme dataset is summarized in Table \ref{tab:datasets}.

\section{Methodology and Experimental Setup}

Our methodology is designed to systematically answer our three research questions using two distinct evaluation settings, as depicted in our overall workflow (Figure \ref{fig:framework}).

Following the population-level template definition (\textit{Related Work}), we adopt a vision-only feature set: template identity is carried by the shared visual form, while overlaid text is the varying content templates are defined against (the local/global decomposition of \citet{dubey2018memesequencer}). This contrasts with tasks like harmful-meme or stance detection, where the text-image relationship itself carries the label \cite{pramanick2021momenta}, and with multimodal humor-understanding work treating a meme's effect as a joint visual-textual reasoning problem \cite{liang2026computational}; template identification instead asks whether an image's \textit{form} recurs, a question the caption's wording does not answer and can actively mislead: near-identical captions recur across different templates, and different captions recur on the same one. Notably, even text-incorporating meme-detection studies rarely ablate its contribution: several models surveyed by \citet{Koutlis2023} add text without isolating its impact, suggesting multimodal superiority for meme-related tasks is itself under-evidenced, not merely bypassed here.

\subsection{Feature Engineering}
To ensure a comprehensive comparison of the related works augmented with our proposed methods, we selected visual features spanning from low-level descriptors to deep semantic embeddings.
\textbf{Low-level descriptors} include RGB and Grayscale Histograms for color/brightness distributions and Local Binary Patterns (LBP) \cite{pietikainen2010local} for textural patterns. \textbf{Content-based fingerprints} build on \citet{zannettou2018origins} and \citet{courtois2023computer}: Perceptual Hashing (pHash) \cite{pHash} for holistic image fingerprinting and ORB \cite{rublee2011orb} for local keypoints robust to scale and rotation. \textbf{Deep semantic embeddings}, following \citet{huertas2024clip}, \citet{bates2023template}, and \citet{pramanick2021momenta}, include CLIP image-encoder embeddings \cite{radford2021learning} for zero-shot semantic capabilities, a fused dual-backbone representation concatenating $\ell_2$-normalized \textbf{SigLIP2} \cite{tschannen2025siglip2multilingualvisionlanguage} and \textbf{DINOv2} \cite{oquab2024dinov2learningrobustvisual} embeddings under mixing weight $\alpha$, and embeddings from the penultimate layer of our fine-tuned DenseNet-121 for task-specific meme features.

Our low-level descriptors follow standard practice in prior meme/IWT classification work \citep{sherratt2023multi, Sharma2020}, enabling fair comparison with that line of work.

\subsection{Models for Comparison}
\label{sec:models}

Meme template identification in a real-world context extends beyond standard closed-set classification: Our analysis of social media data reveals that only about 21\% of social media images belong to well-established templates, the remainder being template-free memes and non-meme images, and the set of \say{known} templates is never complete as meme culture evolves. An effective system must therefore both classify memes from known templates and reliably reject or flag out-of-distribution (OOD) inputs, which our methodology addresses via supervised and unsupervised approaches adapted for semi-open-set recognition.

We stress that our contribution here is a fair, semi-open-set \textit{comparison} of these approaches rather than a set of new classifiers: each method retains its original, published mechanism for template classification, and the only adaptations we introduce are the minimum changes needed to make that mechanism support template-free/non-meme rejection under a common protocol, so that methods designed for closed-set classification can be scored on the same OOD-rejection criterion as those built for it from the outset. Full threshold and hyperparameter selection details, including the two necessary departures from source-paper settings, are given in the Appendix (\textit{Threshold and Hyperparameter Selection}). In the supervised paradigm, this means moving beyond canonical classifiers that assume every test image belongs to a known class: we adapt models to instead flag or reject OOD instances, using, for example, Radius Nearest Neighbors ($r$NN), a distance-based method that designates images without close neighbors as outliers, and a two-stage convolutional neural network (CNN) that first distinguishes template-based memes from other images before proceeding to classification.



\subsubsection{Supervised Approaches}

\paragraph{Multinomial Logistic Regression (MLR)} This model serves as a simple baseline model trained on low-level features (RGB, grayscale, texture features).

\paragraph{Radius Nearest Neighbors ($r$NN)} 
$r$NN classifies images by their neighbors within a fixed radius; query images with no neighbors within that radius are designated outliers (template-free), making the method naturally suited to semi-open-set recognition. We used pHash vectors with Hamming distance and our best-performing CNN's embeddings with cosine similarity, with the radius set to the smallest value at which no labeled sample was classified as template-free.


\paragraph{Transfer Learning and CNNs}

We fine-tuned pretrained ResNet \cite{resnet}, EfficientNetV2 \cite{tan2021efficientnetv2}, and DenseNet \cite{huang2017densely} architectures for meme template recognition. The \textbf{Two-stage DenseNet}, our custom implementation for semi-open-set recognition, adds a binary classification head (meme vs.\ non-meme) before the template classification head.

\paragraph{$r$NN-Feature Matching ($r$NN-FM)} 
$r$NN-FM adapts the work of \citet{courtois2023computer} by combining ORB keypoint matching with network analysis. We replaced the original brute-force matcher with FLANN \cite{suju2017flann} for efficiency. Images are considered a match if they share at least $m$ keypoints within distance $d$. We then construct a distance matrix from these matches and apply the $r$NN algorithm for classification. Optimized parameters for our dataset were $d=27$ and $m=20$.

\paragraph{Sparse Matching (SM)}


SM decouples the meme template (global context) from overlaid content (local context) via sparse representation \cite{dubey2018memesequencer}, using the Sparsity Concentration Index \cite{wright2008robust} to reject template-free images.

\subsubsection{Unsupervised Approaches}

\paragraph{HDBSCAN}
HDBSCAN is applied to CLIP and pHash embeddings, first reduced with UMAP \cite{mcinnes2020umap} to preserve global structure, then clustered with HDBSCAN \cite{mcinnes2017hdbscan} for its noise handling. Clusters are labeled via majority vote over the 80\% labeled ImgFlip portion (with the remaining 20\% held out for evaluation).

\paragraph{Approach of \citet{zannettou2018origins} (DBSCAN on pHash)}



DBSCAN on pHash replicates the pipeline of \citet{zannettou2018origins}. It clusters images using DBSCAN with a Hamming distance metric on pHash fingerprints. We used the authors' recommended threshold ($\delta$=8). Unlike our UMAP+HDBSCAN approach, this method clusters directly on raw hash distances, making it a near-duplicate detection pipeline.

\paragraph{Fused SigLIP2 + DINOv2 Clustering}

Fused  SigLIP2+DINOv2 clustering retains the discovery-then-naming protocol above, replacing pHash and CLIP with complementary frozen vision backbones. SigLIP2 \cite{tschannen2025siglip2multilingualvisionlanguage} provides language-aligned semantic representations, while DINOv2 \cite{oquab2024dinov2learningrobustvisual} captures visual structure such as layout and composition. For each image, we extract the pooled SigLIP2 output and the DINOv2 \texttt{CLS} token at $224$px, $\ell_2$-normalize them, and combine them by weighted concatenation $z(x) = [\sqrt{\alpha}\,h_s(x) \Vert \sqrt{1-\alpha}\,h_d(x)]$, where $\alpha=0.55$ balances semantic and visual cues.

The fused descriptors are reduced to $128$ dimensions with PCA, renormalized, and clustered with HDBSCAN \cite{mcinnes2017hdbscan} using a minimum cluster size of $10$. Each non-noise cluster is represented by its normalized mean, and assignments are refined for one iteration by assigning all discovery images, including HDBSCAN noise, to their nearest centroid under cosine similarity. Clusters are then named by majority vote over the labeled $80\%$ ImgFlip portion, preserving the same naming protocol as the UMAP+HDBSCAN baseline. At test time, images are projected using the fitted PCA and assigned to the nearest centroid, with eligible centroids depending on the setting. In the closed-set setting, clusters that received no labeled votes are discarded, so every prediction is a template name. In the semi-open-set setting, they are retained: an image assigned to a cluster that the labeled ImgFlip portion never reached is labeled template-free, so the clustering's own failure to connect that image to any known template serves as the rejection signal rather than a separate threshold or gate model. Thus, unlike pHash, the representation captures learned semantic and visual structure rather than low-level pixel similarity.

\subsubsection{LLM-Based Approach}

\paragraph{Annotation + Retrieval-Augmented LLM Classification (GPT-5.6 Luna)}

In addition to the trained classifiers and the clustering pipelines described above, we examine whether a general-purpose multimodal language model, GPT-5.6 Luna, can assign templates when its weights are left untouched. However, it is not a purely zero-training baseline: the retrieval step that narrows the candidate set is built from labeled training images, so we describe the approach as frozen-LLM annotation.

Presenting the full inventory of 1,704 canonical templates to the model for every query would be impractical, so we first prune the search space with a lightweight, deterministic retriever. For each image to be classified (the \textit{query}), the retriever ranks all templates by visual similarity to the query. Similarity is a weighted combination of six low-level descriptors computed between the query and training reference images of each template. We draw the reference images and select the weights using only the training portion of the data. The retriever keeps a shortlist of 26 candidate templates, each contributing one training reference image. This fixed budget keeps the visual input compact while retaining the correct template for 98.79\% of held-out ImgFlip queries on average across seeds (Appendix, Table~\ref{tab:llm-retrieval-recall}). We tile the query and the 26 reference images into a single composite image, a \textit{contact sheet}, in which each tile carries a candidate identifier. The model receives the contact sheet and the candidate identifiers with their template labels; it does not see retrieval ranks or scores.

We then apply the model in two stages. In the initial classification stage (V1), \textbf{GPT-5.6 Luna} \cite{openai2026gpt56} selects exactly one candidate from the contact sheet and reports a confidence score. The verification-and-refinement stage (V2) checks every V1 prediction in a separate model call: the verifier sees the query side by side with the reference image of the selected template and decides whether the two show the same underlying template, answering yes or no. If the answer is yes, the prediction is final. If the answer is no, the model re-selects in a second, larger pass over a broader set of retrieved candidates with several reference images per candidate; this pass may also return a no-template decision. For the social media evaluation, a separate meme/non-meme gate precedes template assignment and yields non-meme predictions. The Appendix (\textit{LLM Annotation Protocol}) gives the retriever settings, prompts, and escalation rules.

\subsection{Evaluation Setting 1: Controlled Template Classification (RQ1)}
To answer RQ1, we evaluate model performance in a controlled, closed-set environment using the ImgFlip dataset to study how effectively different methods perform at classifying known meme templates on a large, curated dataset. 

\paragraph{Dataset and Task.} 
Models assign each image to one of 1,704 known template classes. The task is evaluated on ImgFlip test splits containing no template-free or non-meme images, isolating each model's raw template classification capability.

\paragraph{Data Splits and Training.}
Every experiment was repeated over three random seeds, each with an independent stratified 80/20 train/test split to reduce the influence of stochasticity and ensure a fair comparison across methods; each split's test partition contains 30,502 images, and we ensured that every split had samples from all $1{,}704$ templates.

\textbf{Supervised methods} were fit on the 80\% portion, with hyperparameters selected on a stratified 10\% validation subset; the held-out 20\% was never seen during training or selection. Every supervised model was fine-tuned independently for each seed rather than reusing a checkpoint, since roughly $80\%$ of one seed's test set falls within another seed's training portion. 
\textbf{Unsupervised methods} (HDBSCAN, \citet{zannettou2018origins}) clustered a combined corpus comprising the 80\% ImgFlip portion and all 1.1M social media images, except for 1,000 social media images which were sampled in advance for the RQ2 analysis. Of the 1,238,854 images, 17 could not be decoded and were excluded, giving a discovery corpus of $1{,}238{,}837$ (only the ImgFlip portion varies by seed); ImgFlip labels assigned cluster identities via majority vote, and evaluation used the same held-out 20\%.
\textbf{GPT-5.6 Luna} used neither a social media discovery corpus nor fine-tuning. We selected its reference images and retriever weights using only the 80\% ImgFlip training portion of the corresponding split and scored the pipeline on the held-out 20\%, whose labels we used for evaluation only.

\paragraph{Metrics.} 
Performance is measured using Matthews Correlation Coefficient (MCC) \cite{MATTHEWS1975442}, Cohen's Kappa ($\kappa$) \cite{cohen1960coefficient}, and the weighted F1 score, which are well-suited for imbalanced multiclass problems. For each metric, we report the grand mean across seeds with a $95\%$ CI via two-stage bootstrapping (for methodological details, see Appendix, \textit{RQ1 Extended Analysis}).

\subsection{Evaluation Setting 2: Real-World Template Identification (RQ2)}
\paragraph{Dataset and Task.} The evaluation used a manually labeled set of 1,000 images sampled from Reddit, $\mathbb{X}$, and Facebook, requiring models to both classify template-based memes and reject non-template-based content. Methods lacking a native OOD-rejection mechanism (MLR and the single-stage CNNs) are evaluated only in the closed-set RQ1 setting. Our labeling revealed that this sample comprises only 21\% template-based memes, with the remainder being a mix of template-free memes (58.3\%) and non-meme images (20.7\%). This scarcity of established templates matches the recurrence-based definition of templatehood in \textit{Related Work}: most memetic content has not yet accumulated the reuse needed to count as a template.

\paragraph{Ground-Truth Creation and Inter-Annotator Agreement.}
Five annotators who are well acquainted with the ImgFlip template library independently labeled each of the 1,000 images as template-based (while also providing the exact template ID), template-free, or non-meme, applying the criteria defined in the \textit{Defining Memes and Meme Templates} section above. The final label was the majority vote over the five annotators; ties in the votes for five images were resolved through joint discussion among all five annotators. Additional details about the annotation process and a screenshot of the annotation interface (Figure~\ref{fig:annotation-tool}) can be found in the Appendix (\textit{Annotation Interface}).

Inter-annotator agreement was substantial (Fleiss' $\kappa=0.693$ \citep{mchugh2012interrater}; pairwise agreement 81.8\%; individual accuracy vs.\ consensus 89.7\%). Disagreements concentrated at the non-meme vs.\ template-free boundary (64.3\%) and, consistent with the Memetic Images ambiguity discussed in \textit{Related Work}, the template-based vs.\ template-free boundary (32.2\%); template-based memes were rarely confused with non-memes (3.5\%) and received unanimous agreement most often (84.3\% vs.\ 56.9\% for template-free). The dominant ambiguity is thus whether an image is a meme at all, not which template it uses. 

\paragraph{Metrics.}
We use both multiclass metrics (MCC, $\kappa$, weighted F1) to assess classification accuracy and binary metrics (precision, recall, and F1 of the template-based class) to evaluate each model's ability to distinguish template-based memes from anything else, i.e., template-free memes and non-memes. Each reported figure is a point estimate with a $95\%$ CI obtained by bootstrapping the test images (10,000 resamples).

\section{Results}

In this section, we present our findings, structured to directly answer the research questions posed in the Introduction and according to the evaluation described in the previous section.

\subsection{RQ1: Performance in a Controlled Setting}

To answer our first research question, we evaluated the models on the ImgFlip dataset with 1,704 known templates. Table~\ref{tab:imgflip-results} reports the grand mean across the three seeds; the corresponding bootstrap confidence intervals, uniformly narrow and indicating robust results, are given in the Appendix (\textit{RQ1 Extended Analysis}).

\begin{table*}[t]
\centering
\small
\begin{tabular}{l|ll|ccc}
\hline
\textbf{Learning type}        & \multicolumn{1}{l|}{\textbf{Model}}                 & \textbf{Feature}                             & \textbf{F1}    & $\bm{\kappa}$  & \textbf{MCC}   \\ \hline
\multirow{9}{*}{Supervised}   & \multicolumn{1}{l|}{MLR}                            & baseline features                            & 0.971          & 0.969          & 0.969          \\ \cline{2-6} 
                              & \multicolumn{1}{l|}{\multirow{3}{*}{$r$NN}}          & pHash                                        & 0.954          & 0.941          & 0.942          \\
                              & \multicolumn{1}{l|}{}                               & DenseNet embedding                           & 0.973          & 0.969          & 0.969          \\
                              & \multicolumn{1}{l|}{}                               & Feature Matching \cite{courtois2023computer} & 0.916          & 0.882          & 0.886          \\
                              & \multicolumn{1}{l|}{ResNet-18}                      & \multirow{4}{*}{raw image}          & 0.968          & 0.970          & 0.970          \\
                              & \multicolumn{1}{l|}{\textbf{DenseNet-121}}          &                                              & \textbf{0.974} & \textbf{0.974} & \textbf{0.974} \\
                              & \multicolumn{1}{l|}{Two-stage DenseNet-121}         &                                              & 0.944          & 0.928          & 0.929          \\
                              & \multicolumn{1}{l|}{EfficientNetV2}                 &                                              & 0.947          & 0.952          & 0.952          \\ \cline{2-6} 
                              & \multicolumn{2}{l|}{Sparse Matching \cite{dubey2018memesequencer}}                                 & 0.718          & 0.690          & 0.705          \\ \hline

\multirow{4}{*}{Unsupervised} & \multicolumn{1}{l|}{\multirow{2}{*}{UMAP+HDBSCAN}}  & pHash                                        & 0.905          & 0.912          & 0.912      \\
                              & \multicolumn{1}{l|}{}                               & CLIP embedding                               & 0.650          & 0.676          & 0.676          \\ \cline{2-6} 
                              & \multicolumn{1}{l|}{\makecell[l]{DBSCAN-based approach of \\\citet{zannettou2018origins}}}  & pHash                                & 0.676          & 0.728          & 0.728          \\ \cline{2-6} 
                              & \multicolumn{1}{l|}{\makecell[l]{Fused  SigLIP2+DINOv2 \\}} & fused embeddings           & 0.928          & 0.940          & 0.940          \\ \hline
\multirow{1}{*}{LLM}     & \multicolumn{1}{l|}{GPT-5.6 Luna}         & \makecell[l]{retrieval + verification}    & 0.968         & 0.963           & 0.963          \\ \hline 
\end{tabular}
\caption{Performance metrics for models in the controlled setting. Each value is the grand mean obtained via two-stage bootstrapping with 10,000 resamples per seed; confidence intervals are reported in the Appendix, Table~\ref{tab:appendix-rq1-ci}.}
\label{tab:imgflip-results}
\end{table*}

\begin{table*}[t]
\centering
\small
\begin{tabular}{l|l|ccc|cc}
\hline
\multicolumn{1}{c}{\multirow{2}{*}{\textbf{Models}}} & \multicolumn{1}{c|}{\multirow{2}{*}{\textbf{Feature}}} & \multicolumn{3}{c|}{\makecell[l]{\textbf{Overall Performance}\\}} & \multicolumn{2}{c}{\makecell[l]{\textbf{Template-Based vs.} \\ \textbf{Non-Template-Based}}} \\
\multicolumn{1}{c}{} & \multicolumn{1}{c|}{} & \textbf{F1} & $\bm{\kappa}$ & \textbf{MCC} & \textbf{Recall} & \textbf{Prec.} \\ \hline
\multirow{3}{*}{$r$NN} & pHash & 0.556 & 0.254 & 0.266 & 0.590 & 0.449 \\
& \makecell[l]{DenseNet embedding} & 0.600 & 0.362 & 0.379 & 0.857 & 0.468 \\
& \makecell[l]{Feature Matching} & 0.556 & 0.272 & 0.301 & 0.586 & 0.502 \\ \hline
\makecell[l]{Two-stage DenseNet-121} & raw image & 0.199 & 0.149 & 0.235 & 0.876 & 0.272 \\ \hline
\multicolumn{2}{l|}{Sparse Matching \cite{dubey2018memesequencer} } & 0.074 & -0.004 & -0.022 & \textbf{0.981} & 0.209 \\ \hline
\multirow{2}{*}{\makecell[l]{UMAP+HDBSCAN}} & pHash & 0.519 & 0.175 & 0.178 & 0.529 & 0.382 \\
 & CLIP emb. & 0.567 & 0.288 & 0.302 & 0.766 & 0.463 \\ \hline
\makecell[l]{DBSCAN-based approach of \citet{zannettou2018origins}}  & pHash & 0.486 & 0.123 & 0.132 & 0.457 & 0.342 \\ \hline
\makecell[l]{Fused SigLIP2+DINOv2} & \makecell[l]{fused  embeddings} & 0.598 & 0.366 & 0.384 & 0.905 & 0.492 \\ \hline
GPT-5.6 Luna & retrieval + verification & \textbf{0.799} & \textbf{0.656} & \textbf{0.684} & 0.881 & \textbf{0.894} \\ \hline
\end{tabular}
\caption{Performance metrics on the 1,000-image social media subsample. Each value is the point estimate obtained via bootstrapping with 10,000 resamples. The \say{Non-Template-Based} category refers to the union of the template-free plus non-meme images. Confidence intervals and breakdowns for the predicted template-based and true template-based subsets are reported in the Appendix, Table~\ref{tab:appendix-rq2-full}.}
\label{tab:social-media-results}
\end{table*}

DenseNet-121 achieved the strongest supervised performance (MCC 0.974), though the margin over classical methods was small: MLR over color and texture histograms reached MCC 0.969, suggesting deep representations offer limited benefit on clean, well-defined images. The two-stage DenseNet-121 underperformed the single-stage model (MCC 0.929), largely because its binary gate rejected 5.5\% of genuine memes before template classification.

Our fused SigLIP2+DINOv2 pipeline reached MCC 0.940 without using template labels during discovery, close to the strongest supervised methods, while producing substantially fewer, less fragmented clusters than UMAP+HDBSCAN ($7{,}328$ vs.\ $14{,}431$). Representation choice was decisive here too: swapping pHash for CLIP in the same pipeline dropped MCC from 0.912 to 0.676. GPT-5.6 Luna reached MCC 0.963 via retrieval-augmented verification, just below the strongest supervised models. Together, these results suggest performance on this controlled benchmark is driven mainly by visual near-duplicate recognition, with limited additional benefit from more complex learned representations.

\subsection{RQ2: Performance in a Real-World Setting}

To answer our second research question, we assessed the models' robustness and applicability on real-world data using our manually labeled 1,000-image sample of social media content. Table~\ref{tab:social-media-results} reports the point estimates on this sample; the corresponding bootstrap confidence intervals are given in the Appendix (\textit{RQ2 Extended Error Analysis}).

The models show poor overall performance (mean weighted F1 = 0.495, max = 0.799) and, in the binary task of separating template-based memes from everything else, moderate recall (mean = 0.743, max = 0.981) at consistently low precision. The best closed-set MCC (0.974, Table~\ref{tab:imgflip-results}) falls to 0.684, achieved by GPT-5.6 Luna, which also has the highest binary precision (F1 0.799, precision 0.894). Our fused SigLIP2+DINOv2 clustering and $r$NN over DenseNet embeddings are the strongest task-specific methods (MCC 0.384 and 0.379, F1 0.598 and 0.600, respectively). For the task-specific methods, precision rather than recall is the binding constraint: only 210 of the 1,000 images are genuinely template-based, and the best precision achieved by any self-hosted method is 0.502, meaning that roughly half of all template-based claims are incorrect. Luna is the exception, combining a precision of 0.894 with a recall of 0.881 (occupying neither extreme of this trade-off), which suggests the template-based versus non-template-based (template-free plus non-meme) distinction is tractable when a system can reason explicitly about whether a recurring template underlies an image, rather than relying on a fixed geometric acceptance criterion. In the Appendix, we examine what drives this drop: the balance between template-naming accuracy and template-based and template-free rejection, the role of representation, the composition of false positives, and several method-specific failure modes.

\subsection{RQ3: Efficiency and Scalability}
Finally, we consider the practical trade-offs in computational cost. All measurements were taken on an NVIDIA DGX Spark device using one GPU and 18 CPU cores, with the exception of the GPT-5.6 Luna experiments, which used Codex CLI. Cost is governed almost entirely by whether a method encodes each image once or compares images pairwise. The supervised models, the retrieval methods over precomputed representations, and all four clustering pipelines complete in under three hours per seed, and the clustering pipelines do so while processing eight times more data. The two matching-based methods instead compare every query against a reference set of individual images, requiring 20.6 hours per seed for Sparse Matching and 41.1 hours for $r$NN Feature Matching, the latter being roughly 113 times slower per image than our fused SigLIP2+DINOv2 pipeline. GPT-5.6 Luna occupies a third regime bounded by API throughput rather than corpus size, processing the ImgFlip test split at 12,201 images per hour, about 34 times slower than our pipeline, and, unlike every self-hosted custom method here, at a direct per-token monetary cost. Accuracy and efficiency are therefore not in tension among the self-hosted methods: our SigLIP2+DINOv2 pipeline sustains approximately 418,000 images per hour while achieving the strongest unsupervised accuracy, whereas the two slowest self-hosted methods are also among the weakest on real-world data. Detailed timings and further analysis are reported in the Appendix (\textit{RQ3 Extended Efficiency Analysis}).

\section{Conclusion}
This comparative study shows that meme template identification in the wild is a semi-open-set recognition problem, not a standard multiclass task: most real-world content is out-of-distribution (79\%), and the curated-to-wild performance gap is severe (MCC 0.974$\to$0.684), driven almost entirely by the template-based versus template-free decision rather than template naming. Representation matters more than algorithm family: semantic embeddings (our fused SigLIP2+DINOv2 clustering, Luna, $r$NN over DenseNet features) substantially outperform perceptual hashing, which is vulnerable to real-world cropping and recombination. The precision/recall behavior reflects threshold calibration rather than pipeline family: threshold-based hashing is most conservative (recall 0.457), our threshold-free clustering trades precision for recall (0.905/0.492), and Luna escapes the trade-off entirely (0.881/0.894). 

Luna performs best overall but requires a per-image API call, roughly 34 times slower than our fused SigLIP2+DINOv2 pipeline (12,201 vs. 418,000 images/hour) and, unlike self-hosted methods, at a direct monetary cost. Our pipeline therefore offers the strongest self-hosted custom default: it is statistically indistinguishable from $r$NN over DenseNet features on the overall metrics, but recovers considerably more template identity on genuinely template-based images and uses no template labels during discovery at a fraction of the cost. For downstream analyses that only need to name already-identified templates, both methods perform comparably (True Template-Based F1 0.854 and 0.862; Appendix, \textit{RQ2 Extended Error Analysis}), leaving cost and conservativeness as the deciding factors. Luna suits modest-volume settings where the template-based versus template-free decision must be reliable, such as flagging pipelines with costly false positives.

Limitations include a focus on visual templates (excluding video and text-driven formats), a discrepancy between static datasets and evolving live feeds, and a 1,000-image annotated sample that, while reliable in aggregate (Fleiss' $\kappa=0.693$), limits subgroup analysis. Despite these constraints, our framework and released datasets, which we will make publicly available upon acceptance, offer a solid baseline for computational memetics. 

Future work should incorporate text via NLP or VLMs to address \say{Memetic Trends} and rapidly emerging templates \cite{hazman2025beyond}, and use targeted degradation augmentation (cropping, compression, UI elements) to bridge the domain gap between training sets and social media.

\section*{Acknowledgments}
\subsection*{Funding}
D.Á.K. was supported by the Doctoral Students’ Excellence Grant Program (DKÖP-25-1-BME-75), while R.M. was supported by the János Bolyai Research Scholarship of the Hungarian Academy of Sciences. D.Á.K. and R.M. are also supported by the international funding of the 2025 Central Europe Leuven Strategic Alliance (CELSA) project.

\subsection*{LLM use}
GPT-5.6 Luna was used for data filtering and annotation, as described in the main text and in the Appendix. In addition, Claude Sonnet 5 was used to identify and correct spelling, grammatical, and structural errors, as well as to paraphrase selected passages to improve readability and ensure consistency and coherence in the language throughout the study.

{\small
\bibliography{refs}

@article{hazman2025makes,
  author  = {Hazman, M. and McKeever, S. and Griffith, J.},
  title   = {What Makes a Meme a Meme? Identifying Memes for Memetics-Aware Dataset Creation},
  journal = {Proceedings of the International AAAI Conference on Web and Social Media},
  volume  = {19},
  number  = {1},
  pages   = {745--759},
  year    = {2025},
  doi     = {10.1609/icwsm.v19i1.35843},
  url     = {https://doi.org/10.1609/icwsm.v19i1.35843}
}

@article{leach2017social,
  title={The social psychology of the {B}lack {L}ives {M}atter meme and movement},
  author={Leach, Colin Wayne and Allen, Aerielle M},
  journal={Current Directions in Psychological Science},
  volume={26},
  number={6},
  pages={543--547},
  year={2017},
  publisher={Sage Publications Sage CA: Los Angeles, CA}
}

@INPROCEEDINGS{hazman2025beyond,
  author={Hazman, Muzhaffar and McKeever, Susan and Griffith, Josephine},
  booktitle={2025 Fourteenth International Conference on Image Processing, Theory, Tools \& Applications (IPTA)}, 
  title={Beyond Meme Templates: Limitations of Visual Similarity Measures in Meme Matching}, 
  year={2025},
  volume={},
  number={},
  pages={1-6},
  doi={10.1109/IPTA66025.2025.11222070}}

@article{malodia2022meme,
  title={Meme marketing: How can marketers drive better engagement using viral memes?},
  author={Malodia, Suresh and Dhir, Amandeep and Bilgihan, Anil and Sinha, Pranao and Tikoo, Tanishka},
  journal={Psychology \& Marketing},
  volume={39},
  number={9},
  pages={1775--1801},
  year={2022},
  publisher={Wiley Online Library}
}

@article{wang2025internet,
  title={Internet Meme on Social Media: A Comprehensive Review and New Perspectives},
  author={Wang, Bingbing and Lin, Jingjie and Bai, Zhixin and Wang, Zihan and Sun, Shengzhe and Jin, Zhengda and Wen, Zhiyuan and Tu, Geng and Li, Jing and Cambria, Erik and others},
  journal={Information Fusion},
  pages={104102},
  year={2025},
  publisher={Elsevier}
}

@article{shifman2013memes,
  title={Memes in a digital world: Reconciling with a conceptual troublemaker},
  author={Shifman, Limor},
  journal={Journal of Computer-Mediated Communication},
  volume={18},
  number={3},
  pages={362--377},
  year={2013},
  publisher={Oxford University Press Oxford, UK}
}

@inproceedings{dubey2018memesequencer,
  title={Memesequencer: Sparse matching for embedding image macros},
  author={Dubey, Abhimanyu and Moro, Esteban and Cebrian, Manuel and Rahwan, Iyad},
  booktitle={Proceedings of the 2018 World Wide Web Conference},
  pages={1225--1235},
  year={2018}
}

@article{courtois2023computer,
  title={Computer Vision and Internet Meme Genealogy: An Evaluation of Image Feature Matching as a Technique for Pattern Detection},
  author={Courtois, Cedric and Frissen, Thomas},
  journal={Communication Methods and Measures},
  volume={17},
  number={1},
  pages={17--39},
  year={2023},
  publisher={Taylor \& Francis}
}

@inproceedings{zannettou2018origins,
  title={On the origins of memes by means of fringe web communities},
  author={Zannettou, Savvas and Caulfield, Tristan and Blackburn, Jeremy and De Cristofaro, Emiliano and Sirivianos, Michael and Stringhini, Gianluca and Suarez-Tangil, Guillermo},
  booktitle={Proceedings of the Internet Measurement Conference 2018},
  pages={188--202},
  year={2018}
}

@inproceedings{pramanick2021momenta,
    title = "{MOMENTA}: A Multimodal Framework for Detecting Harmful Memes and Their Targets",
    author = "Pramanick, Shraman  and
      Sharma, Shivam  and
      Dimitrov, Dimitar  and
      Akhtar, Md. Shad  and
      Nakov, Preslav  and
      Chakraborty, Tanmoy",
    editor = "Moens, Marie-Francine  and
      Huang, Xuanjing  and
      Specia, Lucia  and
      Yih, Scott Wen-tau",
    booktitle = "Findings of the Association for Computational Linguistics: EMNLP 2021",
    month = nov,
    year = "2021",
    address = "Punta Cana, Dominican Republic",
    publisher = "Association for Computational Linguistics",
    url = "https://aclanthology.org/2021.findings-emnlp.379/",
    doi = "10.18653/v1/2021.findings-emnlp.379",
    pages = "4439--4455"
}

@inproceedings{radford2021learning,
  title={Learning transferable visual models from natural language supervision},
  author={Radford, Alec and Kim, Jong Wook and Hallacy, Chris and Ramesh, Aditya and Goh, Gabriel and Agarwal, Sandhini and Sastry, Girish and Askell, Amanda and Mishkin, Pamela and Clark, Jack and others},
  booktitle={International Conference on Machine Learning},
  pages={8748--8763},
  year={2021},
  organization={PMLR}
}

@inproceedings{rublee2011orb,
  title={ORB: An efficient alternative to SIFT or SURF},
  author={Rublee, Ethan and Rabaud, Vincent and Konolige, Kurt and Bradski, Gary},
  booktitle={2011 International Conference on Computer Vision},
  pages={2564--2571},
  year={2011},
  organization={Ieee}
}

@inproceedings{theisen2023motif,
  title={Motif mining: Finding and summarizing remixed image content},
  author={Theisen, William and Cedre, Daniel Gonzalez and Carmichael, Zachariah and Moreira, Daniel and Weninger, Tim and Scheirer, Walter},
  booktitle={Proceedings of the IEEE/CVF Winter Conference on Applications of Computer Vision},
  pages={1319--1328},
  year={2023}
}

@article{davison2012language,
  title={The language of internet memes},
  author={Davison, Patrick},
  journal={The Social Media Reader},
  pages={120--134},
  year={2012}
}

@article{resnet,
  title={Deep Residual Learning for Image Recognition},
  author={Kaiming He and X. Zhang and Shaoqing Ren and Jian Sun},
  journal={2016 IEEE Conference on Computer Vision and Pattern Recognition (CVPR)},
  year={2015},
  pages={770-778},
  url={https://api.semanticscholar.org/CorpusID:206594692}
}

@inproceedings{huang2017densely,
  title={Densely connected convolutional networks},
  author={Huang, Gao and Liu, Zhuang and Van Der Maaten, Laurens and Weinberger, Kilian Q},
  booktitle={Proceedings of the IEEE Conference on Computer Vision and Pattern Recognition},
  pages={4700--4708},
  year={2017}
}

@article{pietikainen2010local,
  title={Local binary patterns},
  author={Pietik{\"a}inen, Matti},
  journal={Scholarpedia},
  volume={5},
  number={3},
  pages={9775},
  year={2010}
}

@article{mcinnes2017hdbscan,
  title={hdbscan: Hierarchical density based clustering.},
  author={McInnes, Leland and Healy, John and Astels, Steve},
  journal={Journal of Open Source Software},
  volume={2},
  number={11},
  pages={205},
  year={2017}
}

@misc{pHash,
  author = {Evan Klinger and David Starkweather},
  title = {pHash, The open source perceptual hash library},
  howpublished = {https://www.phash.org},
    year= {2008},
  note = {Accessed: 2024-04-25}
}

@article{mcinnes2020umap,
  author    = {McInnes, Leland and Healy, John and Saul, Nathaniel and Großberger, Lukas},
  title     = {UMAP: Uniform Manifold Approximation and Projection},
  journal   = {Journal of Open Source Software},
  year      = {2018},
  volume    = {3},
  number    = {29},
  pages     = {861},
  publisher = {The Open Journal},
  doi       = {10.21105/joss.00861},
  url       = {https://doi.org/10.21105/joss.00861}
}

@article{MATTHEWS1975442,
title = {Comparison of the predicted and observed secondary structure of T4 phage lysozyme},
journal = {Biochimica et Biophysica Acta (BBA) - Protein Structure},
volume = {405},
number = {2},
pages = {442-451},
year = {1975},
issn = {0005-2795},
doi = {https://doi.org/10.1016/0005-2795(75)90109-9},
url = {https://www.sciencedirect.com/science/article/pii/0005279575901099},
author = {B.W. Matthews}
}

@article{cohen1960coefficient,
  title={A coefficient of agreement for nominal scales},
  author={Cohen, Jacob},
  journal={Educational and Psychological Measurement},
  volume={20},
  number={1},
  pages={37--46},
  year={1960},
  publisher={Sage Publications Sage CA: Thousand Oaks, CA}
}

@inproceedings{tan2021efficientnetv2,
  title={Efficientnetv2: Smaller models and faster training},
  author={Tan, Mingxing and Le, Quoc},
  booktitle={International conference on machine learning},
  pages={10096--10106},
  year={2021},
  organization={PMLR}
}

@inproceedings{sherratt2023multi,
  title={Multi-channel Convolutional Neural Network for Precise Meme Classification},
  author={Sherratt, Victoria and Pimbblet, Kevin and Dethlefs, Nina},
  booktitle={Proceedings of the 2023 ACM International Conference on Multimedia Retrieval},
  pages={190--198},
  year={2023}
}

@inproceedings{deka2017rico,
  title={Rico: A mobile app dataset for building data-driven design applications},
  author={Deka, Biplab and Huang, Zifeng and Franzen, Chad and Hibschman, Joshua and Afergan, Daniel and Li, Yang and Nichols, Jeffrey and Kumar, Ranjitha},
  booktitle={Proceedings of the 30th Annual ACM Symposium on User Interface Software and Technology},
  pages={845--854},
  year={2017}
}

@inproceedings{plummer2015flickr30k,
  title={Flickr30k entities: Collecting region-to-phrase correspondences for richer image-to-sentence models},
  author={Plummer, Bryan A and Wang, Liwei and Cervantes, Chris M and Caicedo, Juan C and Hockenmaier, Julia and Lazebnik, Svetlana},
  booktitle={Proceedings of the IEEE International Conference on Computer Vision},
  pages={2641--2649},
  year={2015}
}

@inproceedings{suju2017flann,
  title={FLANN: Fast approximate nearest neighbour search algorithm for elucidating human-wildlife conflicts in forest areas},
  author={Suju, D Arul and Jose, Hancy},
  booktitle={2017 Fourth International Conference on Signal Processing, Communication and Networking (ICSCN)},
  pages={1--6},
  year={2017},
  organization={IEEE}
}

@article{wright2008robust,
  title={Robust face recognition via sparse representation},
  author={Wright, John and Yang, Allen Y and Ganesh, Arvind and Sastry, S Shankar and Ma, Yi},
  journal={IEEE Transactions on Pattern Analysis and Machine Intelligence},
  volume={31},
  number={2},
  pages={210--227},
  year={2008},
  publisher={IEEE}
}

@article{mchugh2012interrater,
  title={Interrater reliability: the kappa statistic},
  author={McHugh, Mary L},
  journal={Biochemia medica},
  volume={22},
  number={3},
  pages={276--282},
  year={2012},
  publisher={Medicinska naklada}
}

@inproceedings{tommasini2023imkg,
  title={IMKG: The internet meme knowledge graph},
  author={Tommasini, Riccardo and Ilievski, Filip and Wijesiriwardene, Thilini},
  booktitle={European Semantic Web Conference},
  pages={354--371},
  year={2023},
  organization={Springer}
}

@inproceedings{joshi2024contextualizing,
  title={Contextualizing Internet Memes Across Social Media Platforms},
  author={Joshi, Saurav and Ilievski, Filip and Luceri, Luca},
  booktitle={Companion Proceedings of the ACM on Web Conference 2024},
  pages={1831--1840},
  year={2024}
}

@inproceedings{bates2023template,
    title = "A Template Is All You Meme",
    author = "Bates, Luke  and
      Christensen, Peter Ebert  and
      Nakov, Preslav  and
      Gurevych, Iryna",
    editor = "Chiruzzo, Luis  and
      Ritter, Alan  and
      Wang, Lu",
    booktitle = "Proceedings of the 2025 Conference of the Nations of the Americas Chapter of the Association for Computational Linguistics: Human Language Technologies (Volume 1: Long Papers)",
    month = apr,
    year = "2025",
    address = "Albuquerque, New Mexico",
    publisher = "Association for Computational Linguistics",
    url = "https://aclanthology.org/2025.naacl-long.525/",
    doi = "10.18653/v1/2025.naacl-long.525",
    pages = "10443--10475",
    ISBN = "979-8-89176-189-6"
}

@inproceedings{huertas2024clip,
  title={A CLIP-based siamese approach for meme classification},
  author={Huertas-Tato, Javier and Koutlis, Christos and Papadopoulos, Symeon and Camacho, David and Kompatsiaris, Ioannis},
  booktitle={2024 International Joint Conference on Neural Networks (IJCNN)},
  pages={1--8},
  year={2024},
  organization={IEEE}
}

@misc{tschannen2025siglip2multilingualvisionlanguage,
      title={SigLIP 2: Multilingual Vision-Language Encoders with Improved Semantic Understanding, Localization, and Dense Features}, 
      author={Michael Tschannen and Alexey Gritsenko and Xiao Wang and Muhammad Ferjad Naeem and Ibrahim Alabdulmohsin and Nikhil Parthasarathy and Talfan Evans and Lucas Beyer and Ye Xia and Basil Mustafa and Olivier Hénaff and Jeremiah Harmsen and Andreas Steiner and Xiaohua Zhai},
      year={2025},
      eprint={2502.14786},
      archivePrefix={arXiv},
      primaryClass={cs.CV},
      url={https://arxiv.org/abs/2502.14786}, 
}

@misc{oquab2024dinov2learningrobustvisual,
      title={DINOv2: Learning Robust Visual Features without Supervision}, 
      author={Maxime Oquab and Timothée Darcet and Théo Moutakanni and Huy Vo and Marc Szafraniec and Vasil Khalidov and Pierre Fernandez and Daniel Haziza and Francisco Massa and Alaaeldin El-Nouby and Mahmoud Assran and Nicolas Ballas and Wojciech Galuba and Russell Howes and Po-Yao Huang and Shang-Wen Li and Ishan Misra and Michael Rabbat and Vasu Sharma and Gabriel Synnaeve and Hu Xu and Hervé Jegou and Julien Mairal and Patrick Labatut and Armand Joulin and Piotr Bojanowski},
      year={2024},
      eprint={2304.07193},
      archivePrefix={arXiv},
      primaryClass={cs.CV},
      url={https://arxiv.org/abs/2304.07193}, 
}

@misc{liang2026computational,
      title={Computational Humor with Multimodal LLMs: Methods, Datasets, Evaluation, and Challenges}, 
      author={Tuo Liang and Zhe Hu and Disheng Liu and Jing Li and Yu Yin},
      year={2026},
      eprint={2607.19011},
      archivePrefix={arXiv},
      primaryClass={cs.CL},
      url={https://arxiv.org/abs/2607.19011}, 
}

@book{Milner2016,
    author = {Milner, Ryan M.},
    title = {The World Made Meme: Public Conversations and Participatory Media},
    publisher = {The MIT Press},
    year = {2016},
    month = {10},
    isbn = {9780262034999},
    doi = {10.7551/mitpress/9780262034999.001.0001},
    url = {https://doi.org/10.7551/mitpress/9780262034999.001.0001},
}

@conference{Sharma2020,
author={Chhavi Sharma and Viswanath Pulabaigari},
title={A Curious Case of Meme Detection: An Investigative Study},
booktitle={Proceedings of the 16th International Conference on Web Information Systems and Technologies - Volume 1: WEBIST,},
year={2020},
pages={327-338},
publisher={SciTePress},
organization={INSTICC},
doi={10.5220/0010110203270338},
isbn={978-989-758-478-7},
}

@ARTICLE{Koutlis2023,
  title    = "{MemeTector}: enforcing deep focus for meme detection",
  author   = "Koutlis, Christos and Schinas, Manos and Papadopoulos, Symeon",
  journal  = "International Journal of Multimedia Information Retrieval",
  volume   =  12,
  number   =  1,
  pages    = "11",
  month    =  may,
  year     =  2023
}

@article{Du2020,
author = {Du, Yuhao and Masood, Muhammad and Joseph, Kenneth},
year = {2020},
month = {05},
pages = {153-164},
title = {Understanding Visual Memes: An Empirical Analysis of Text Superimposed on Memes Shared on Twitter},
volume = {14},
journal = {Proceedings of the International AAAI Conference on Web and Social Media},
doi = {10.1609/icwsm.v14i1.7287}
}

@article{Beskow2020,
title = {The evolution of political memes: Detecting and characterizing internet memes with multi-modal deep learning},
journal = {Information Processing \& Management},
volume = {57},
number = {2},
pages = {102170},
year = {2020},
issn = {0306-4573},
doi = {https://doi.org/10.1016/j.ipm.2019.102170},
url = {https://www.sciencedirect.com/science/article/pii/S0306457319307988},
author = {David M. Beskow and Sumeet Kumar and Kathleen M. Carley}
}

@conference{SharmaPulaBaigariDas2020,
author={Chhavi Sharma and Viswanath Pulabaigari and Amitava Das},
title={Meme vs. Non-meme Classification using Visuo-linguistic Association},
booktitle={Proceedings of the 16th International Conference on Web Information Systems and Technologies - WEBIST},
year={2020},
pages={353-360},
publisher={SciTePress},
organization={INSTICC},
doi={10.5220/0010176303530360},
isbn={978-989-758-478-7},
issn={2184-3252},
}

@ARTICLE{luceri2026coordinated,
  title     = "Coordinated inauthentic behavior on {TikTok}: Challenges and
               opportunities for detection in a video-first ecosystem",
  author    = "Luceri, Luca and Salkar, Tanishq Vijay and Balasubramanian,
               Ashwin and Pinto, Gabriela and Sun, Chenning and Ferrara, Emilio",
  journal   = "Proceedings of the International AAAI Conference on Web and
               Social Media",
  publisher = "Association for the Advancement of Artificial Intelligence
               (AAAI)",
  volume    =  20,
  number    =  1,
  pages     = "1533--1550",
  month     =  may,
  year      =  2026
}

@misc{qu2022disinfomeme,
      title={DisinfoMeme: A Multimodal Dataset for Detecting Meme Intentionally Spreading Out Disinformation}, 
      author={Jingnong Qu and Liunian Harold Li and Jieyu Zhao and Sunipa Dev and Kai-Wei Chang},
      year={2022},
      eprint={2205.12617},
      archivePrefix={arXiv},
      primaryClass={cs.CL},
      url={https://arxiv.org/abs/2205.12617}, 
}

@misc{openai2026gpt56,
    author  = {{OpenAI}},
    title   = {{GPT-5.6}: Frontier Intelligence That Scales with Your Ambition},
    year    = {2026},
    month   = jul,
    url     = {https://openai.com/index/gpt-5-6/},
    urldate = {2026-09-02}
  }

@misc{openai2026codexloop,
    author    = {Bolin, Michael},
    title     = {Unrolling the Codex Agent Loop},
    publisher = {OpenAI},
    year      = {2026},
    month     = jan,
    url       = {https://openai.com/index/unrolling-the-codex-agent-loop/},
    urldate   = {2026-09-02}
  }

@article{lememematch2026,
  author  = {Le, D. T. A. and K{\"o}ller, D. {\'A}kos and Deng, Q. and Molontay, R.},
  title   = {MemeMatch: A Large-Scale Dual-Context Multimodal Dataset and Retrieval System for Internet Memes},
  journal = {Proceedings of the International AAAI Conference on Web and Social Media},
  volume  = {20},
  number  = {1},
  pages   = {2828--2838},
  year    = {2026},
  doi     = {10.1609/icwsm.v20i1.42785},
  url     = {https://doi.org/10.1609/icwsm.v20i1.42785}
}

@article{sherratt2026clustering,
  title     = {Clustering Internet Memes with Metric Learning and Dynamic Modality Weighting},
  author    = {Sherratt, Victoria and Dethlefs, Nina},
  journal   = {Proceedings of the International AAAI Conference on Web and Social Media},
  volume    = {20},
  number    = {1},
  pages     = {2131--2148},
  year      = {2026},
  doi       = {10.1609/icwsm.v20i1.42741}
}
}

\appendix
\section{Appendix}

\subsection{Template Example}

\begin{figure}[!ht]
\centering
\includegraphics[width=\columnwidth]{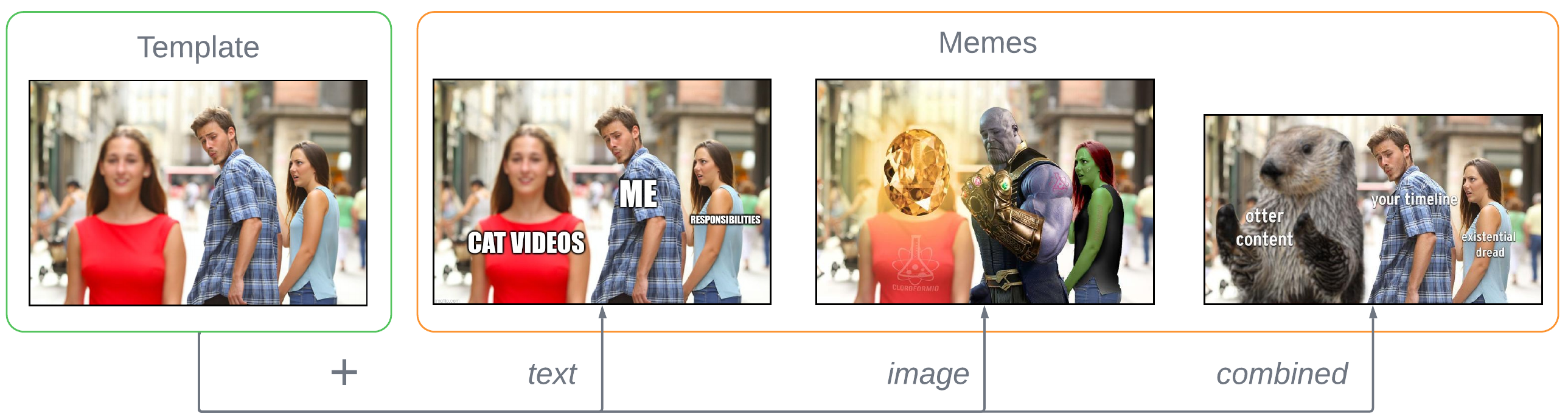}
\caption{Different realizations of the ``Distracted Boyfriend'' template.}
\label{fig:template}
\end{figure}

Figure \ref{fig:template} shows multiple independently created instances of the \say{Distracted Boyfriend} template, illustrating how a single recognizable visual form supports a wide range of edits: some instances differ only in the overlaid text; others substitute different imagery while preserving the template's layout and character poses; and still others combine both. This flexibility is central to what makes a format reusable at scale: the template supplies a stable, recognizable structure, while each instance's specific substitution carries the joke.

\subsection{ImgFlip Deduplication}
\label{appendix:dedup}
We screened the collection for duplicate identities in two complementary ways. First, we embedded the reference image of every template with CLIP \cite{radford2021learning} and ranked all pairs by cosine similarity, treating the highest-scoring pairs as merge candidates. Second, since embedding similarity alone misses duplicates that differ in cropping, resolution, or overlaid text, we asked GPT-5.6 Luna \cite{openai2026gpt56} to flag further candidates from side-by-side visual comparison and from recurring label confusions observed in a preliminary classification pass on an earlier, unmerged version of the dataset. This procedure is distinct from the final experiments reported in Results, and used here only to surface merge candidates. Neither procedure merged anything on its own: the authors inspected every proposed group and decided individually whether the labels referred to the same template.

\subsection{Threshold and Hyperparameter Selection}
\label{appendix:thresholds}

Wherever a source paper specifies a value, we retain it. The pHash-DBSCAN pipeline of \citet{zannettou2018origins} is run with their published 64-bit perceptual hash, Hamming radius $\delta=8$, and $\mathit{minPts}=5$; the feature-matching method of \citet{courtois2023computer} retains their descriptor-distance threshold $d=27$ (see Supervised Approaches for our adaptation of $m$). Two departures from source-paper settings are unavoidable, and we state them explicitly.

First, sparse matching \citep{dubey2018memesequencer} does not specify a sparsity regularization strength, so $\alpha$ must be chosen; we also downsample to $64\times64$ pixels rather than the original $48\times48$. Our label space is considerably larger than that in the original setting (1,704 templates with up to 30 references each, yielding 40,773 dictionary atoms), and at $48\times48$ the dictionary would be $17.7\times$ overcomplete, compared with $10.0\times$ at $64\times64$; because $\ell_1$ recovery degrades as this ratio grows, the higher resolution keeps the problem comparably conditioned at our scale.

Second, and more fundamentally, a rejection threshold cannot be inherited from a closed-set paper, since the original method had no rejection step to parameterize. Every such threshold is therefore introduced by us. For $r$NN we select the neighborhood radius as the smallest value at which no labeled validation sample is classified as an outlier, evaluated on a held-out split carved from the training portion, never on the test set. For the two-stage classifier, we select the stage-one operating point on the same validation split, maximizing the geometric mean of sensitivity and specificity. For the unsupervised methods, no threshold is tuned at all: a cluster is named by majority vote over its labeled members, and clusters without labeled members are rejected. Beyond these rejection parameters, we perform no hyperparameter search for the self-hosted methods; the CNNs and the unsupervised pipelines run at fixed settings, with validation used only for early stopping. The one exception is the retriever of the LLM pipeline: we select its six similarity weights from nine predefined configurations using training images only (see \textit{LLM Annotation Protocol}).

\subsection{LLM Annotation Protocol}
\label{appendix:llm-protocol}

\paragraph{Candidate retrieval.}
The retriever compares each query with training reference images using six low-level descriptors. We resize images to a maximum side length of 640 pixels and extract (i) $16\times16$ grayscale intensities, (ii) 63 low-frequency discrete cosine transform (DCT) coefficients (the upper-left $8\times8$ block of the DCT of a $32\times32$ grayscale image, excluding the DC term), (iii) $16\times16$ edge maps, (iv) a $12\times4$ hue--saturation histogram, (v) the aspect ratio, and (vi) a 63-bit perceptual hash obtained by median-thresholding the coefficients in (ii). We mean-center and $\ell_2$-normalize the grayscale and edge vectors, $\ell_2$-normalize the DCT vectors without mean-centering, and sum-normalize, square-root transform, and $\ell_2$-normalize the color histogram. We compare grayscale, DCT, and edge descriptors by cosine similarity rescaled to $[0,1]$, color histograms by cosine similarity without rescaling, hashes by the fraction of agreeing bits, and aspect ratios $a_q$ and $a_r$ by $\exp(-|\log(a_q/a_r)|)$. To reduce the influence of caption bars above or below an image, we compare each query in four views: the full image, and the image with the top 18\%, the bottom 18\%, or 12\% from both top and bottom removed. References use their full-image view. For each descriptor we keep the highest similarity across query views, and we combine the six similarities by a weighted sum.

\paragraph{Reference banks and shortlist construction.}
For each ImgFlip split, we select all reference images deterministically from its 80\% training portion. A \emph{main bank} stores three training exemplars per template and a \emph{supplementary bank} stores one fixed training reference (template without texts) per template; each bank yields its own ranking, which we call a retrieval branch. A template's main bank score is that of its highest-scoring exemplar. The shortlist consists of the 24 top-ranked templates from the main bank plus the two top-ranked templates from the supplementary bank that are not already selected, giving 26 distinct candidates. On the contact sheet, main bank candidates appear with their highest-scoring exemplar and supplementary candidates with their fixed reference, so the model sees exactly one reference image per candidate.

\paragraph{Weight selection.}
For each split, we choose the six similarity weights from nine predefined configurations, evaluated on training-only tuning images disjoint from the reference banks. During tuning, each descriptor's similarity is maximized over the three main-bank exemplars before the weighted sum is computed. At inference, the weighted sum is computed separately for each exemplar, and the highest exemplar score determines the template's main-bank score. The selected weights are shared by both retrieval branches. We select the configuration that maximizes Recall@24, with Recall@10 and Recall@1 as successive tie-breakers, and freeze it before test-set retrieval and LLM inference. In descriptor order (grayscale, DCT, edges, color, aspect ratio, hash), the selected weights are $(0.27, 0.15, 0.12, 0.30, 0.08, 0.08)$ for seed 40 and $(0.409, 0, 0.273, 0.205, 0.113, 0)$ for seeds 41 and 42; the released code includes the full grid.

\paragraph{Candidate coverage.}
We measure candidate-set recall as the fraction of queries whose ground-truth template appears among the retrieved candidates, before any LLM decision. The 26-candidate shortlist reaches a mean recall of 98.79\% across the three held-out ImgFlip splits (Table~\ref{tab:llm-retrieval-recall}); expanding the set to 130 candidates raises coverage to 99.33\%, a gain of 0.54 percentage points for five times as many candidates. These figures support a compact initial shortlist with the broad fallback reserved for cases remaining unresolved after the focused reassessment and verification. We fixed the budget by design and did not tune it by a candidate-count ablation; the test-set measurements describe its coverage and played no role in selecting it.

\begin{table}[t]
\centering
\small
\begin{tabular}{lrrr}
\hline
\textbf{Split} & \textbf{R@1} & \textbf{R@26} & \textbf{R@130} \\
\hline
Seed 40 & 95.36 & 98.65 & 99.24 \\
Seed 41 & 95.33 & 99.08 & 99.54 \\
Seed 42 & 95.33 & 98.65 & 99.22 \\
\hline
Mean    & 95.34 & 98.79 & 99.33 \\
\hline
\end{tabular}
\caption{Candidate-set recall (\%) of the retriever on the 30,502 held-out ImgFlip images of each seed. R@1 uses the main retrieval branch; R@26 and R@130 use the 24+2 and 120+10 candidate unions rather than a single global ranking. R@130 is measured over all test queries, although the broad fallback runs only for cases remaining unresolved after the focused reassessment and verification. These are retrieval-coverage measures, not final classification scores.}
\label{tab:llm-retrieval-recall}
\end{table}

\paragraph{Initial classification (V1).}
We process queries in batches of ten. Each query appears on its own contact sheet: the query image above a four-column grid of its 26 candidates in shuffled order. The prompt lists the candidate identifiers and template labels and withholds retrieval ranks and scores, filenames, source paths, and gold labels. GPT-5.6 Luna answers under a fixed output schema with exactly one candidate identifier and a confidence value from 0 to 100; this value is a model self-assessment and is not calibrated.

\paragraph{Verification and refinement (V2).}
A fresh model call (the verifier) checks every V1 prediction: it sees the query side by side with the selected reference image, without the template label or the V1 confidence, and answers whether the two share the same underlying template. Accepted predictions retain the V1 label. Rejected predictions enter a focused reassessment over the 12 highest-ranked candidates from the main retrieval branch, with up to four training references each, and the verifier checks any concrete choice in the same way. Choices rejected at this point, and \texttt{NO\_TEMPLATE} answers, proceed to a broad fallback over 130 candidates (120 main-branch candidates plus 10 additional labels from the supplementary branch), which returns a template label or
\texttt{NO\_TEMPLATE}.

\paragraph{Social media extension.}
 The 1,000-image social media sample uses the seed-40 training references. A separate Luna classifier judges from the visible content whether each image is a meme. The V1--V2 template pipeline is run on all 1,000 images; during output consolidation, a negative meme decision overrides its predictions with \texttt{NOT\_A\_MEME}. For images judged to be memes, concrete predictions from the broad fallback undergo an additional blind verification against up to four training references of the predicted template; rejection yields   \texttt{NO\_TEMPLATE}. This label means only that the pipeline accepted no match in the inventory; the image may still use a recurring template outside it.

\paragraph{Inference configuration.}
All calls use GPT-5.6 Luna at low reasoning effort with structured JSON output, orchestrated through the Codex CLI \cite{openai2026codexloop}. V1 and its initial verifier process batches of up to ten queries; the focused reassessment, its verifier, and the broad search process one query at a time. The social-media meme classifier and the additional broad-prediction verifier also use batches of up to ten. We pass images under anonymized identifiers, disable tool use and network access, and exclude gold labels from every prompt.

\subsection{Annotation Interface}

\begin{figure*}[t]
    \centering
    \includegraphics[trim={3cm 0cm 1cm 0cm},clip,width=0.82\linewidth]{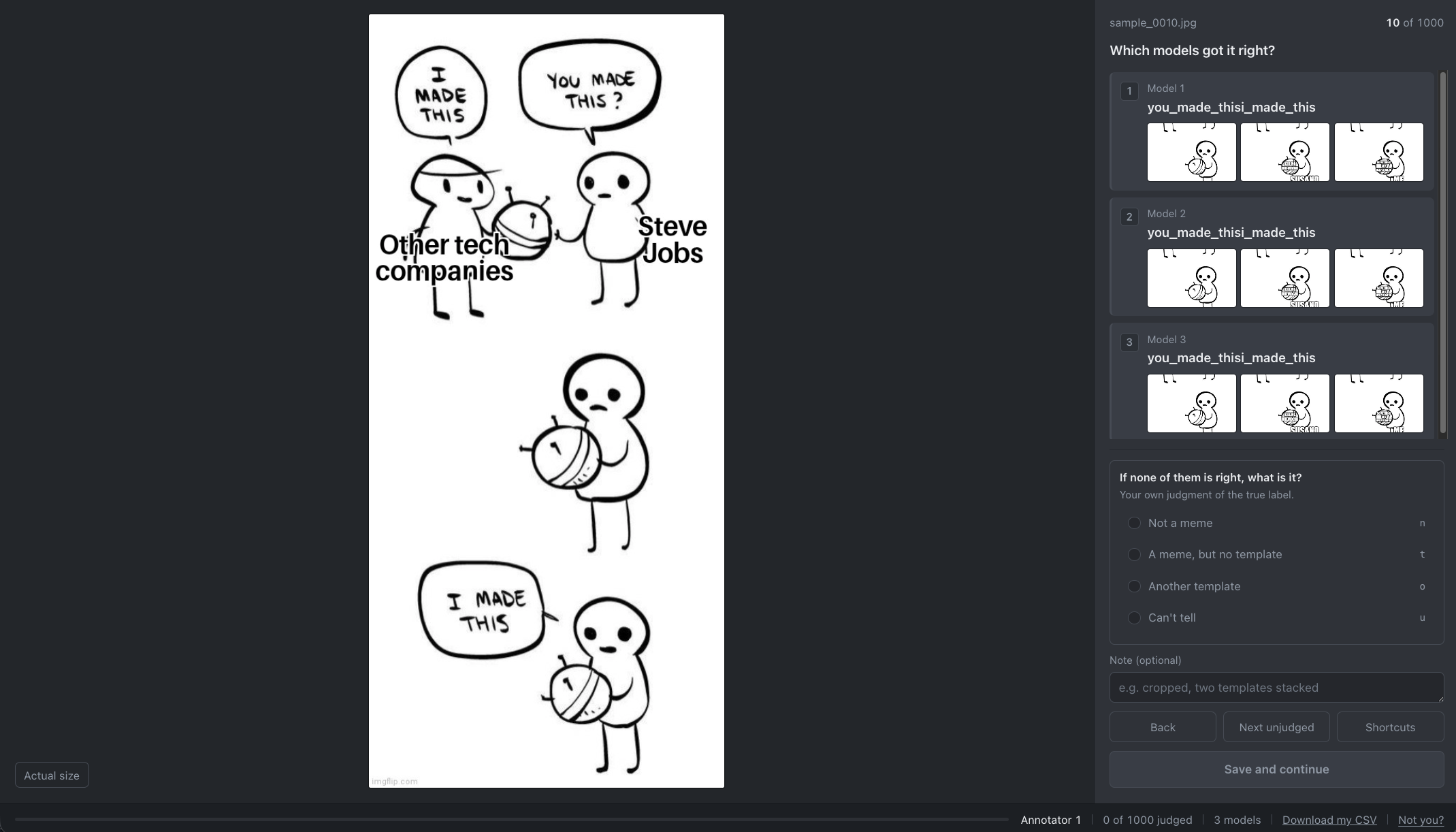}
    \caption{The annotation interface. Each image is shown alongside the three systems' predictions (anonymized, order-shuffled, and confidence-free, so annotators cannot favor a familiar method) and the annotator either checks whichever prediction is correct or, if
    none is, supplies their own judgment of the true label.}
    \label{fig:annotation-tool}
\end{figure*}

Figure~\ref{fig:annotation-tool} shows the interface used to collect the ground-truth; the \say{Can't tell} option is recorded separately and mapped to template-free during consolidation. Because the label space spans 1,704 templates---far more than an annotator can be expected to recall by name---and each template requires a separate lookup to confirm its ID, the interface surfaces one prediction per image from each of three independently developed methods (Two-stage DenseNet-121 classifier, GPT-5.6 Luna, and our fused  SigLIP2+DINOv2 pipeline), letting annotators check off any prediction they judge correct rather than searching the full inventory from scratch; checking a prediction adopts it directly as the recorded true label. Annotators may check more than one prediction, but checking a single one is sufficient to proceed; if the checked predictions disagree with one another, the interface displays a warning asking the annotator to keep only the one that matches, though it does not block saving, so the first checked prediction in display order is recorded if the warning is not acted on. Predictions are shown only as \say{Model 1}, \say{Model 2}, \say{Model 3}, with display order shuffled independently for every image, identically for all annotators, and confidence scores withheld entirely, so annotators cannot learn which model produced which suggestion from its identity, position, or reporting pattern. Where a prediction names a specific template, up to three example images of that template are shown alongside it, so the annotator can confirm or reject the claim visually rather than judging it from the template name alone; where a method instead predicts non-meme or template-free, no example images are shown for that slot, and the prediction serves only to indicate that the corresponding model found no template match.

If none of the three predictions is correct, the annotator instead records their own judgment directly: not a meme, a meme with no template, a specific template found via search, or can't tell. The \say{Another template} search opens a box over the 1,704 template names in the label space: results are ranked on every keystroke by a blend of whole-word overlap, character trigrams, and prefix matching, which together tolerate partial recall and typing errors, and the five closest matches are shown with their own example images for confirmation. A typed name matching a known template after normalizing case, spacing, and punctuation is snapped to its canonical spelling before being recorded, so the same template never appears under multiple spellings in the data. Where the annotator is confident that an image is built on a template outside our 1,704-template inventory, no canonical match exists to snap to, and the typed name is instead recorded verbatim, preserving it for future analysis; such judgments are counted as template-based during consolidation, since the annotator has asserted that a template underlies the image even though its name lies outside our inventory. In practice, they rarely survive to the final label: of the 1,124 template-based judgments cast, 139 were typed by hand rather than adopted from a prediction, and 96 of those named a template outside the inventory, yet almost all were minority votes, leaving only 2 of the 210 consolidated template-based images with a label our inventory does not contain.

Surfacing model predictions in this way necessarily risks anchoring annotators toward candidates the models themselves proposed, even when incorrect. We mitigate this in three ways: predictions are shown only as anonymized, shuffled, confidence-free candidates rather than attributed to a named, potentially trusted method; every template candidate requires visual confirmation against example images before it can be accepted rather than being judged on the model's say-so alone; and whenever the annotator is making their own judgment, the template-free, non-meme, and \say{Another template} options are presented as equally prominent, unranked alternatives rather than as a fallback. We nonetheless cannot rule out residual anchoring bias toward the models' collective blind spots, since a template none of the three methods proposed is, by construction, harder for an annotator to surface unaided, and we note this as a limitation of our ground-truth procedure.

\subsection{RQ1 Extended Analysis}
\label{app:rq1-uncertainty}

Each method is trained and evaluated independently on three seeds. For each seed, we draw $B = 10{,}000$ bootstrap resamples (with replacement) of that seed's test images and compute the metric on each, giving a within-seed bootstrap mean $\bar\theta_r$ and bootstrap variance $v_r$; the latter estimates the test-sampling variance conditional on the model trained on that seed. We then apply the law of total variance to separate a between-seed (training and partition) component, $V_{\text{between}}$, the sample variance of the three $\bar\theta_r$ around the grand mean $\bar\theta = \frac{1}{3}\sum_r \bar\theta_r$, from a within-seed (test-sampling) component, $V_{\text{within}}$, the average of the three $v_r$. Table~\ref{tab:appendix-rq1-ci} reports $\bar\theta$ with a $95\%$ confidence interval on this mean estimate, $\bar\theta \pm t_{2,\,0.975} \cdot \mathrm{SE}(\bar\theta)$, where $\mathrm{SE}(\bar\theta) = \sqrt{V_{\text{between}}/3 + V_{\text{within}}/(3B)}$ and the $t$-distribution with $R - 1 = 2$ degrees of freedom accounts for the between-seed component being estimated from only three splits. With only two degrees of freedom, this interval is markedly wider than a normal approximation would give, reflecting genuine uncertainty about between-seed variance rather than a looser standard. 

The intervals separate the table into broad tiers but not into a strict ordering: the four strongest methods on MCC (DenseNet-121, ResNet-18, MLR, and $r$NN over DenseNet embeddings) overlap, while the gap to perceptual hashing and to the matching baselines is far larger than any interval. The widest bands belong to EfficientNetV2-S and the two-stage CNN, whose run-to-run variation is roughly four times the median across methods.

\begin{table*}[t]
\centering
\small
\begin{tabular}{ll|ccc}
\hline
\multicolumn{1}{c}{\multirow{2}{*}{\textbf{Models}}} & \multicolumn{1}{c|}{\multirow{2}{*}{\textbf{Feature}}} & \multicolumn{3}{c}{\makecell[l]{\textbf{Closed-Set Performance}}} \\
\multicolumn{1}{c}{} & \multicolumn{1}{c|}{} & \textbf{F1} & $\bm{\kappa}$ & \textbf{MCC} \\ \hline
\multicolumn{2}{l|}{MLR (color and texture histograms)} & \makecell{0.971\\{\scriptsize [0.970, 0.972]}} & \makecell{0.969\\{\scriptsize [0.968, 0.970]}} & \makecell{0.969\\{\scriptsize [0.968, 0.970]}} \\ \hline
$\bm{r}$NN & pHash & \makecell{0.954\\{\scriptsize [0.952, 0.956]}} & \makecell{0.941\\{\scriptsize [0.938, 0.944]}} & \makecell{0.942\\{\scriptsize [0.939, 0.944]}} \\
$r$NN & \makecell[l]{DenseNet \\ embedding} & \makecell{0.973\\{\scriptsize [0.971, 0.975]}} & \makecell{0.969\\{\scriptsize [0.964, 0.973]}} & \makecell{0.969\\{\scriptsize [0.964, 0.973]}} \\
$r$NN & \makecell[l]{Feature \\ Matching} & \makecell{0.916\\{\scriptsize [0.913, 0.919]}} & \makecell{0.882\\{\scriptsize [0.877, 0.886]}} & \makecell{0.886\\{\scriptsize [0.882, 0.890]}} \\ \hline
ResNet-18 & raw image & \makecell{0.968\\{\scriptsize [0.966, 0.970]}} & \makecell{0.970\\{\scriptsize [0.967, 0.974]}} & \makecell{0.970\\{\scriptsize [0.967, 0.974]}} \\
\makecell[l]{DenseNet-121} & raw image & \makecell{\textbf{0.974}\\{\scriptsize [0.969, 0.979]}} & \makecell{\textbf{0.974}\\{\scriptsize [0.969, 0.980]}} & \makecell{\textbf{0.974}\\{\scriptsize [0.969, 0.980]}} \\
\makecell[l]{Two-stage \\ DenseNet-121} & raw image & \makecell{0.944\\{\scriptsize [0.934, 0.955]}} & \makecell{0.928\\{\scriptsize [0.912, 0.943]}} & \makecell{0.929\\{\scriptsize [0.914, 0.944]}} \\ \hline
EfficientNetV2-S & raw image & \makecell{0.947\\{\scriptsize [0.929, 0.964]}} & \makecell{0.952\\{\scriptsize [0.936, 0.969]}} & \makecell{0.952\\{\scriptsize [0.936, 0.969]}} \\ \hline
\multicolumn{2}{l|}{Sparse Matching} & \makecell{0.718\\{\scriptsize [0.712, 0.723]}} & \makecell{0.690\\{\scriptsize [0.684, 0.696]}} & \makecell{0.705\\{\scriptsize [0.699, 0.710]}} \\ \hline
\makecell[l]{UMAP+\\HDBSCAN} & pHash & \makecell{0.905\\{\scriptsize [0.902, 0.908]}} & \makecell{0.912\\{\scriptsize [0.909, 0.915]}} & \makecell{0.912\\{\scriptsize [0.909, 0.915]}} \\
\makecell[l]{UMAP+\\HDBSCAN} & CLIP emb. & \makecell{0.650\\{\scriptsize [0.641, 0.660]}} & \makecell{0.676\\{\scriptsize [0.672, 0.680]}} & \makecell{0.676\\{\scriptsize [0.672, 0.680]}} \\ \hline
\multicolumn{2}{l|}{Approach of \citet{zannettou2018origins}:} & \multirow{2}{*}{\makecell{0.676\\{\scriptsize [0.668, 0.683]}}} & \multirow{2}{*}{\makecell{0.728\\{\scriptsize [0.722, 0.734]}}} & \multirow{2}{*}{\makecell{0.728\\{\scriptsize [0.722, 0.734]}}} \\
DBSCAN & pHash & & & \\ \hline
\makecell[l]{Fused \\ SigLIP2+DINOv2} & \makecell[l]{fused embeddings} & \makecell{0.928\\{\scriptsize [0.924, 0.933]}} & \makecell{0.940\\{\scriptsize [0.937, 0.944]}} & \makecell{0.940\\{\scriptsize [0.937, 0.944]}} \\ \hline
\multicolumn{1}{l}{GPT-5.6 Luna} & \makecell[l]{raw image + \\template inventory} & \makecell{0.968\\{\scriptsize [0.963, 0.973]}} & \makecell{0.963\\{\scriptsize [0.957, 0.969]}} & \makecell{0.963\\{\scriptsize [0.957, 0.969]}} \\
\hline
\end{tabular}
\caption{RQ1 closed-set performance with confidence intervals on the mean. Within each seed, $10{,}000$ bootstrap resamples of the test images estimate the within-seed (test-sampling) variance; the three seed-level bootstrap means additionally give a between-seed (training/partition) variance. The reported $95\%$ interval is $\bar\theta \pm t_{2,0.975}\cdot\mathrm{SE}(\bar\theta)$, combining both components via the law of total variance. Bold marks the highest value in each column.}
\label{tab:appendix-rq1-ci}
\end{table*}

\subsection{RQ2 Extended Error Analysis}
\label{appendix:rq2-failure-modes}

Table \ref{tab:appendix-rq2-full} reports the full RQ2 breakdown, including error bars and the multiclass accuracy restricted to the predicted template-based and true template-based subsets omitted from the main-text summary (Table \ref{tab:social-media-results}).

\begin{table*}[t]
\centering
\scriptsize
\setlength{\tabcolsep}{2.5pt}
\begin{tabular}{ll|ccc|ccc|ccc|cc}
\hline
\multicolumn{1}{c}{\multirow{2}{*}{\textbf{Models}}} & \multicolumn{1}{c|}{\multirow{2}{*}{\textbf{Feature}}} & \multicolumn{3}{c|}{\makecell[l]{\textbf{Overall Performance}}} & \multicolumn{3}{c|}{\makecell[l]{\textbf{Performance on} \\ \textbf{Predicted Template-Based}}} & \multicolumn{3}{c|}{\makecell[l]{\textbf{Performance on}\\ \textbf{True Template-Based}}} & \multicolumn{2}{c}{\makecell[l]{\textbf{Template-Based vs. } \\\textbf{Non-Template-Based} } } \\
\multicolumn{1}{c}{} & \multicolumn{1}{c|}{} & \textbf{F1} & $\bm{\kappa}$ & \textbf{MCC} & \textbf{F1} & $\bm{\kappa}$ & \textbf{MCC} & \textbf{F1} & $\bm{\kappa}$ & \textbf{MCC} & \textbf{Recall} & \textbf{Prec.} \\ \hline
$\bm{r}$NN & pHash & \makecell{.556\\{\tiny [.522,.590]}} & \makecell{.254\\{\tiny [.206,.301]}} & \makecell{.266\\{\tiny [.216,.314]}} & \makecell{.364\\{\tiny [.306,.422]}} & \makecell{.370\\{\tiny [.314,.427]}} & \makecell{.406\\{\tiny [.350,.461]}} & \makecell{.504\\{\tiny [.436,.573]}} & \makecell{.487\\{\tiny [.418,.554]}} & \makecell{.530\\{\tiny [.467,.591]}} & \makecell{.590\\{\tiny [.523,.657]}} & \makecell{.449\\{\tiny [.391,.507]}} \\
$r$NN & \makecell[l]{DenseNet \\ embedding} & \makecell{.600\\{\tiny [.566,.633]}} & \makecell{.362\\{\tiny [.316,.407]}} & \makecell{.379\\{\tiny [.332,.424]}} & \makecell{.414\\{\tiny [.366,.465]}} & \makecell{.422\\{\tiny [.373,.472]}} & \makecell{.461\\{\tiny [.414,.508]}} & \makecell{.789\\{\tiny [.733,.842]}} & \makecell{.773\\{\tiny [.715,.827]}} & \makecell{.782\\{\tiny [.727,.833]}} & \makecell{.857\\{\tiny [.809,.902]}} & \makecell{.468\\{\tiny [.418,.518]}} \\
$r$NN & \makecell[l]{Feature \\ Matching} & \makecell{.556\\{\tiny [.521,.590]}} & \makecell{.272\\{\tiny [.225,.320]}} & \makecell{.301\\{\tiny [.251,.350]}} & \makecell{.403\\{\tiny [.341,.464]}} & \makecell{.405\\{\tiny [.344,.467]}} & \makecell{.451\\{\tiny [.392,.508]}} & \makecell{.495\\{\tiny [.426,.564]}} & \makecell{.473\\{\tiny [.404,.541]}} & \makecell{.523\\{\tiny [.460,.583]}} & \makecell{.586\\{\tiny [.520,.652]}} & \makecell{.502\\{\tiny [.438,.565]}} \\ \hline
\makecell[l]{Two-stage \\ DenseNet-121} & raw image & \makecell{.199\\{\tiny [.173,.225]}} & \makecell{.149\\{\tiny [.124,.174]}} & \makecell{.235\\{\tiny [.201,.269]}} & \makecell{.216\\{\tiny [.185,.248]}} & \makecell{.227\\{\tiny [.196,.259]}} & \makecell{.297\\{\tiny [.263,.330]}} & \makecell{.742\\{\tiny [.682,.800]}} & \makecell{.730\\{\tiny [.669,.789]}} & \makecell{.738\\{\tiny [.679,.794]}} & \makecell{.876\\{\tiny [.829,.919]}} & \makecell{.272\\{\tiny [.239,.306]}} \\ \hline
\makecell[l]{Sparse \\ Matching} & raw image & \makecell{.074\\{\tiny [.059,.092]}} & \makecell{-.004\\{\tiny [-.010,.001]}} & \makecell{-.022\\{\tiny [-.053,.009]}} & \makecell{.067\\{\tiny [.052,.083]}} & \makecell{.068\\{\tiny [.053,.084]}} & \makecell{.091\\{\tiny [.072,.111]}} & \makecell{.331\\{\tiny [.269,.396]}} & \makecell{.316\\{\tiny [.256,.378]}} & \makecell{.326\\{\tiny [.265,.388]}} & \makecell{\textbf{.981}\\{\tiny [.961,.996]}} & \makecell{.209\\{\tiny [.184,.235]}} \\ \hline
\makecell[l]{UMAP+\\HDBSCAN} & pHash & \makecell{.519\\{\tiny [.487,.552]}} & \makecell{.175\\{\tiny [.126,.225]}} & \makecell{.178\\{\tiny [.129,.229]}} & \makecell{.314\\{\tiny [.260,.369]}} & \makecell{.318\\{\tiny [.263,.372]}} & \makecell{.364\\{\tiny [.310,.417]}} & \makecell{.461\\{\tiny [.393,.529]}} & \makecell{.439\\{\tiny [.373,.507]}} & \makecell{.483\\{\tiny [.420,.545]}} & \makecell{.529\\{\tiny [.462,.596]}} & \makecell{.382\\{\tiny [.325,.439]}} \\
\makecell[l]{UMAP+\\HDBSCAN} & CLIP emb. & \makecell{.567\\{\tiny [.534,.601]}} & \makecell{.288\\{\tiny [.240,.337]}} & \makecell{.302\\{\tiny [.253,.351]}} & \makecell{.303\\{\tiny [.255,.353]}} & \makecell{.314\\{\tiny [.266,.363]}} & \makecell{.356\\{\tiny [.307,.405]}} & \makecell{.538\\{\tiny [.468,.607]}} & \makecell{.519\\{\tiny [.451,.588]}} & \makecell{.535\\{\tiny [.468,.602]}} & \makecell{.766\\{\tiny [.707,.823]}} & \makecell{.463\\{\tiny [.410,.515]}} \\ \hline
\makecell[l]{DBSCAN \\ \citeauthor{zannettou2018origins} \\ (\citeyear{zannettou2018origins})} & pHash & \makecell{.486\\{\tiny [.453,.520]}} & \makecell{.123\\{\tiny [.077,.171]}} & \makecell{.132\\{\tiny [.082,.182]}} & \makecell{.202\\{\tiny [.155,.250]}} & \makecell{.205\\{\tiny [.159,.252]}} & \makecell{.240\\{\tiny [.190,.291]}} & \makecell{.283\\{\tiny [.222,.345]}} & \makecell{.274\\{\tiny [.214,.334]}} & \makecell{.322\\{\tiny [.260,.383]}} & \makecell{.457\\{\tiny [.388,.525]}} & \makecell{.342\\{\tiny [.286,.398]}} \\ \hline
\makecell[l]{Fused \\ SigLIP2+DINOv2 \\ (ours)} & \makecell[l]{fused \\ embeddings} & \makecell{.598\\{\tiny [.568,.629]}} & \makecell{.366\\{\tiny [.321,.411]}} & \makecell{.384\\{\tiny [.338,.429]}} & \makecell{.452\\{\tiny [.403,.502]}} & \makecell{.457\\{\tiny [.408,.506]}} & \makecell{.508\\{\tiny [.464,.552]}} & \makecell{.854\\{\tiny [.804,.899]}} & \makecell{.841\\{\tiny [.789,.888]}} & \makecell{.844\\{\tiny [.794,.890]}} & \makecell{.905\\{\tiny [.863,.942]}} & \makecell{.492\\{\tiny [.443,.542]}} \\ \hline
GPT-5.6 Luna & \makecell[l]{retrieval + \\verification} & \makecell{\textbf{.799}\\{\tiny [.770,.827]}} & \makecell{\textbf{.656}\\{\tiny [.613,.700]}} & \makecell{\textbf{.684}\\{\tiny [.645,.723]}} & \makecell{\textbf{.863}\\{\tiny [.813,.908]}} & \makecell{\textbf{.858}\\{\tiny [.808,.905]}} & \makecell{\textbf{.864}\\{\tiny [.817,.907]}} & \makecell{\textbf{.862}\\{\tiny [.814,.906]}} & \makecell{\textbf{.846}\\{\tiny [.797,.893]}} & \makecell{\textbf{.853}\\{\tiny [.808,.896]}} & \makecell{.881\\{\tiny [.836,.922]}} & \makecell{\textbf{.894}\\{\tiny [.850,.934]}} \\
\hline
\end{tabular}
\caption{Full breakdown of RQ2 performance. The “Non-Template-Based” category refers to the union of the template-free plus non-meme images. Each cell reports the mean over $10{,}000$ bootstrap resamples of the benchmark images and the corresponding $95\%$ percentile interval, including the Predicted Template-Based and True Template-Based multiclass subsets omitted from the main-text summary. Leading zeros are omitted. Bold marks the best value in each column.}
\label{tab:appendix-rq2-full}
\end{table*}

\textbf{The drop is concentrated in rejection, not naming.} On the 210 images whose consensus label is a specific template, our fused SigLIP2+DINOv2 pipeline and GPT-5.6 Luna name the correct template for 177 and 178 of them (84.3\% and 84.8\% accuracy), yet their overall MCC of 0.384 and 0.684 falls far short of the 0.940 and 0.963 they reach on ImgFlip, since overall RQ2 metrics also penalize incorrectly accepting or rejecting non-template-based content, a source of error the closed-set RQ1 setting cannot produce, as every closed-set image is guaranteed to belong to some template. This indicates that the drop relative to RQ1 is driven mainly by the template-based versus non-template-based decision rather than by template naming itself, suggesting future progress depends less on finer-grained template inventories than on a more precise operational account of what makes an image template-based.

\textbf{Representation matters more than algorithm family.} Across the table, semantic-embedding methods (fused SigLIP2+DINOv2, Luna, and $r$NN over DenseNet features) achieve True Template-Based F1 scores between 0.789 and 0.862, substantially outperforming perceptual-hashing methods (0.283 to 0.504), since a single perceptual hash is highly sensitive to the cropping, recombination, and compression typical of real-world reposts, whereas semantic embeddings remain comparatively stable. The one partial exception is UMAP+HDBSCAN over CLIP embeddings, which, despite also using a semantic representation, falls between the two regimes (True Template-based F1 0.538), plausibly because CLIP's global, language-aligned pooling discards the layout-level detail that our fused SigLIP2+DINOv2 representation retains and that template identity depends on.

\textbf{Over-prediction dominates the error budget.} Because only 210 of the 1,000 images are genuinely template-based, precision is the binding constraint for the self-hosted methods. The false positives consist overwhelmingly of template-free memes rather than non-memes: 164 of 196 for our own pipeline, indicating that these models separate memes from non-memes reasonably well but struggle with the template-based versus one-off distinction, which is also a boundary our annotators found difficult (32.2\% of disagreements, second only to the non-meme versus template-free boundary).

\textbf{Degenerate recall.} Sparse Matching attains the nominally highest recall (0.981) by labeling 984 of the 1,000 images as template-based. Its precision (0.209) is indistinguishable from the 21.0\% base rate, and its overall MCC (-0.022) is the lowest in the table. The high recall therefore reflects an absence of discrimination rather than genuine sensitivity and should not be interpreted as strong performance.

\textbf{Cost of a two-stage gate.} The Two-stage DenseNet-121 exhibits the same pathology in a milder form. Its binary gate fires on 676 images, of which 428 are template-free, yielding a precision of 0.272 and the second-lowest overall F1 (0.199), below that of the single-stage DenseNet-121 on the template-based block. Separating rejection from classification does not, on this data, improve either component.

\textbf{Failure in both directions.} The approach of \citet{zannettou2018origins} (DBSCAN on pHash) is the weakest member of the clustering family, combining the lowest recall (0.457) with one of the lowest precision scores (0.342) and naming only 58 templates correctly (27.6\%). Its 8-bit Hamming radius admits only near-pristine reposts, so most genuine template instances fall outside every cluster, while the clusters that do form are contaminated by visually similar but unrelated images. This behavior appears to be a structural property of hashing at social-media scale rather than an artifact of parameter selection.

\textbf{Comparison with Du et al. (2020).} The results contrast with those of \citet{Du2020}, whose clustering pipeline (\say{ClusterMatching}) on a related task showed the opposite asymmetry (recall 0.09, precision 0.65). The divergence reflects matching strictness, not pipeline family: their KYM-derived inventory rejects any insufficiently close match, systematically missing memes outside it, while our threshold-free nearest-cluster assignment over a 1.1M-image discovery corpus recovers more true instances at the cost of admitting template-free ones. The same ordering holds within our results, with threshold-based methods occupying the low-recall region (0.457 for \citet{zannettou2018origins}, 0.590 for $r$NN+pHash) and threshold-free assignment the high-recall region (0.905, ours), indicating the precision-recall asymmetry reflects acceptance-threshold placement rather than an intrinsic property of clustering. GPT-5.6 Luna is the strongest counter-evidence to any intrinsic trade-off, combining recall 0.881 with precision 0.894: a system reasoning explicitly about template recurrence need not sacrifice one for the other, suggesting the asymmetries above stem from using a fixed geometric criterion as a proxy for a semantic judgment. The operative design question is therefore not whether to cluster, but how conservatively to calibrate the rejection criterion.

\subsection{RQ3 Extended Efficiency Analysis}
\label{appendix:rq3-efficiency}

Table~\ref{tab:appendix-rq3-timing} reports the cost of a complete run for each method, comprising representation extraction, model fitting or clustering, and evaluation. The presented values are wall-clock means over the three seeds.

\begin{table*}[t]
\centering
\small
\begin{tabular}{ll|rrr}
\hline
\textbf{Model} & \textbf{Feature} & \textbf{Images} & \textbf{Hours} & \textbf{Img/hour} \\ \hline
ResNet-18 & raw image & 152,509 & 0.43 & 353,000 \\
DenseNet-121 & raw image & 152,509 & 0.62 & 245,000 \\
EfficientNetV2-S & raw image & 152,509 & 0.74 & 206,000 \\
Two-stage DenseNet-121 & raw image & 152,509 & 2.29 & 67,000 \\ \hline
MLR & color+texture & 152,509 & 2.36 & 65,000 \\ \hline
$r$NN & pHash & 152,509 & 0.24 & 638,000 \\
$r$NN & DenseNet emb. & 152,509 & 0.29 & 531,000 \\
$r$NN & Feature Matching & 152,509 & 41.12 & 3,700 \\ \hline
Sparse Matching & raw image & 152,509 & 20.57 & 7,400 \\ \hline
DBSCAN \citep{zannettou2018origins} & pHash & $1{,}238{,}837$ & 0.40 & 3,080,000 \\
UMAP+HDBSCAN & pHash & 1,238,837 & 1.98 & 627,000 \\
UMAP+HDBSCAN & CLIP embedding & 1,238,837 & 1.63 & 763,000 \\
Fused  SigLIP2+DINOv2 & fused embeddings & 1,238,837 & 2.97 & 418,000 \\ \hline
GPT-5.6 Luna & retrieval+verification & 30,502 & 2.50$^\dagger$ & 12,201 \\ \hline
\end{tabular}
\caption{Wall-clock cost per seed, averaged over three seeds on a single machine. The clustering pipelines process ImgFlip together with 1.1M unlabeled social media images, hence the larger image count. Throughput is images processed per hour of wall-clock time. $^\dagger$GPT-5.6 Luna's figure is the mean of two seeds (2h39m32s, 2h20m47s); the third run (approx.\ 3h25m) was affected by repeated machine instability and is excluded from the average, though it is directionally consistent. Unlike the other rows, this figure covers only V1+V2 test-time inference on the 30,502-image evaluation split; one-time retrieval-prototype construction from the training portion is not included.}
\label{tab:appendix-rq3-timing}
\end{table*}

\textbf{Two cost regimes.} Methods linear in corpus size encode each image once and work in a reduced vector space (65,000–3.1M images/hour, regardless of paradigm). The superlinear methods, $r$NN Feature Matching and Sparse Matching, compare every query against a reference set individually, so cost grows with the product of query and reference sizes, resulting in being two orders of magnitude slower (3,700 and 7,400 images/hour), with the gap widening for larger corpora.

\textbf{A third, API-bound regime.} GPT-5.6 Luna does not fit either category above: its retrieval-narrowed pipeline queries only a handful of candidates per image rather than encoding a corpus or comparing against every reference, yet it processes the 30,502-image ImgFlip test split at just 12,201 images per hour under 30-way parallelism. This is roughly 34 times slower than our fused  SigLIP2+DINOv2 pipeline despite this narrower per-query scope, since its cost is bounded by API request latency and provider-side throughput rather than local compute. It is also the only method in this comparison with a direct monetary cost: averaged across seeds, each run consumed approximately 252 million input and 2.73 million output tokens ($\approx$8,300 input / 90 output tokens per image), a cost the self-hosted methods do not incur at all. This makes Luna's practicality a function of API budget and rate limits rather than hardware, and its cost would scale roughly linearly with corpus size, unlike the sublinear-per-image cost of an already-encoded local pipeline.

\textbf{Implications for corpus scale.} Extrapolating from measured throughput, ten million images would take about 24 hours for our pipeline but roughly 112 days for $r$NN Feature Matching. This means that at social-media scale, matching methods are viable only on curated subsets, restricting them to analyzing known templates rather than discovering them.

\textbf{Where the supervised cost lies.} Single-stage models are cheap since only the head is trained on a frozen backbone (0.43–0.74 hours); the two-stage variant costs more (2.29 hours) by training two networks without improving accuracy. MLR (2.36 hours) is comparable, but nearly all its cost is CPU-side computation of the 1,052-dimensional descriptors, not fitting.

\textbf{Throughput is not a proxy for quality.} The fastest method, DBSCAN over pHash at 3.1 million images per hour, is also the weakest of the clustering family (MCC 0.728 closed set, 0.132 real-world). Its speed reflects how little work an 8-bit Hamming radius performs. Our fused SigLIP2+DINOv2 pipeline is roughly seven times slower per image but recovers substantially more templates, which is the trade-off worth making at these absolute cost levels.

\end{document}